\documentclass[twocolumn,floatfix]{revtex4-2}
\pdfoutput=1
\usepackage{graphicx}
\usepackage{dcolumn}
\usepackage{longtable}
\usepackage{amsmath}
\usepackage{amssymb}
\usepackage{color}
\usepackage{isotope}

\begin{document}
\title{Parameter-free deformation variables of the proxy-SU(3) symmetry in even-even atomic nuclei with ${\bf Z=28-82}$, ${\bf N=28-126}$}

\author
{Dennis Bonatsos$^1$, V. K. B. Kota$^2$, Andriana Martinou$^1$,  S. K. Peroulis$^1$, D.~Petrellis$^3$, P. Vasileiou$^4$, T. J. Mertzimekis$^5$, and N. Minkov$^6$ }

\affiliation
{$^1$Institute of Nuclear and Particle Physics, National Centre for Scientific Research ``Demokritos'', GR-15310 Aghia Paraskevi, Attiki, Greece}

\affiliation
{$^2$ Physical Research Laboratory, Ahmedabad 380 009, India} 

\affiliation
{$^3$ Physics Department, Aristotle University of Thessaloniki, Thessaloniki GR-54124, Greece}

\affiliation
{$^4$ Horia Hulubei National Institute for R{\&}D in Physics and Nuclear Engineering, Strada Reactorului 30, POB MG6, RO-077125
Bucharest-M\v{a}gurele, Romania}

\affiliation
{$^5$  Department of Physics, National and Kapodistrian University of Athens, Zografou Campus, GR-15784 Athens, Greece}

\affiliation
{$^6$Institute of Nuclear Research and Nuclear Energy, Bulgarian Academy of Sciences, 72 Tzarigrad Road, 1784 Sofia, Bulgaria}

\begin{abstract}

The proxy-SU(3) approximation to the shell model, which restores the SU(3) symmetry of the 3-dimensional harmonic oscillator beyond the $sd$ shell, predicts the collective deformation variables $\beta$ and $\gamma$ of even-even atomic nuclei in a parameter-free way, based on the most symmetric irreducible representation (irrep) of SU(3) allowed by the Pauli principle and the short-range nature of the nucleon-nucleon interaction, which in group theoretical language is the highest weight (hw) irrep. In the few cases in which the hw irrep turns out to be completely symmetric, thus being able to accommodate only the ground state band, the next hw (nhw) irrep becomes indispensable. In the present article complete tables of the hw and nhw irreps are given for all atomic nuclei ranging from $Z=28$, $N=28$ to $Z=82$, $N=126$, along with the corresponding parameter-free predictions for the deformation variables $\beta$ and $\gamma$. A few examples using the tabulated results for providing microscopic insight for specific effects in various regions of the nuclear chart are also given.

\end{abstract}

\maketitle

\section{Introduction}

Symmetries have been used in nuclear physics for a long time, since the introduction of the SU(4) symmetry by Wigner in 1937 \cite{Wigner1937}. Wigner has shared the Physics Nobel Prize in 1963 \cite{Nobel1972} with Mayer and Jensen, who introduced in 1949 \cite{Mayer1948,Mayer1949,Haxel1949,Mayer1955} the nuclear shell model \cite{Heyde1990,Talmi1993}, still remaining the standard microscopic model of nuclear structure. The nuclear shell model is based on the three-dimensional isotropic harmonic oscillator (3D-HO) \cite{Wybourne1974,Moshinsky1996,Iachello2006}, to which the spin-orbit force is added \cite{Mayer1948,Mayer1949,Haxel1949,Mayer1955} in order to reproduce the experimentally observed nuclear magic numbers, i.e. the proton and neutron numbers at which nuclei exhibit increased stability. The various shells of the 3D-HO are known to be characterized by overall U(N) symmetries possessing SU(3) subalgebras \cite{Bonatsos1986}. In 1958 Elliott 
\cite{Elliott1958a,Elliott1958b,Elliott1963,Elliott1968,Harvey1968} realized that the SU(3) subalgebra of the U(6) symmetry characterizing the nuclear $sd$ shell is related to the nuclear deformation, thus bridging the gap between the microscopic nuclear shell model and the macroscopic nuclear collective model, introduced by Bohr and Mottelson in 1952 \cite{Bohr1952,Bohr1953,Bohr1998a,Bohr1998b}, in which nuclear properties are described in terms of the collective deformation variables $\beta$ and $\gamma$, related to the deviation from sphericity and the deviation from triaxiality, respectively.  

Since Elliott's seminal work, SU(3) symmetry has been widely used in nuclear structure \cite{Kota2020}. In the shell model framework, in which it is known that the SU(3) symmetry of the 3D-HO is broken beyond the $sd$ shell by the spin-orbit interaction, several approximations restoring the SU(3) symmetry in heavier shells have been developed, including the pseudo-SU(3) symmetry \cite{Arima1969,Hecht1969,RatnaRaju1973,Draayer1982,Draayer1983,Draayer1984,Bahri1992,Ginocchio1997}, the quasi-SU(3) symmetry \cite{Zuker1995,Zuker2015}, and the proxy-SU(3) symmetry \cite{Bonatsos2017a,Bonatsos2017b,Bonatsos2023}. More recently, symmetry-adapted no-core shell model calculations \cite{Dytrych2008,Launey2015,Launey2016,Launey2020,Launey2021} have become affordable, taking advantage of the symplectic symmetry \cite{Rosensteel1977,Rosensteel1980,Rowe1985} in no-core shell-model calculations \cite{Navratil2000a,Navratil2000b}, albeit only in light nuclei up to now. On the other hand, Monte Carlo \cite{Honma1995,Honma1996,Mizusaki1996} shell model calculations \cite{Otsuka1998,Otsuka2001,Shimizu2001,Otsuka2022} have  recently reached heavy nuclei in the rare earth region \cite{Otsuka2025}. 

An alternative path has been taken in 1975 by Arima and Iachello \cite{Arima1975}, through the introduction of the Interacting Boson Model (IBM) 
\cite{Arima1976,Arima1978,Arima1979,Iachello1987,Casten1988,Iachello1991,Casten1993,Talmi1993,Frank2005}, in which correlated valence fermion pairs are treated as bosons. On one hand, the use of bosons offers a tremendous reduction of the computational burden, allowing for fast calculations of spectra and electromagnetic transition rates 
\cite{Scholten1991,Otsuka1985,Casperson2012}. On the other hand, the symmetries occurring in the IBM, U(5) for nearly spherical vibrational nuclei \cite{Arima1976}, SU(3) for well deformed axially symmetric nuclei \cite{Arima1978}, and O(6) for nuclei with shapes soft towards triaxiality \cite{Arima1979}, have to be broken in order to reach agreement to the data. Despite this difficulty, the IBM has also been very useful for the description of shape/phase transitions 
\cite{Scholten1978,Feng1981,Iachello1998,Jolie2001,Jolie2002,Warner2002,Bonatsos2024a} at various places on the nuclear chart, related to critical point symmetries \cite{Iachello2000,Iachello2001,Casten2006,Casten2007,Casten2009,Cejnar2009,Cejnar2010} introduced in the framework of the collective nuclear model. 

The present work is focused on the proxy-SU(3) approximation to the shell model, introduced in 2017 \cite{Bonatsos2017a,Bonatsos2017b} and initially justified 
\cite{Bonatsos2017a} within the Nilsson model \cite{Nilsson1955,Nilsson1995}, which is a modified version of the 3D-HO model allowing for axially symmetric deformations, prolate (rugby-ball-like) or oblate (pancake-like). The proxy-SU(3) approximation 
has later been connected \cite{Martinou2020,Bonatsos2020b} to the shell model, by employing a relatively simple unitary transformation \cite{Martinou2020}. It should be noticed that the pseudo-SU(3) approximation to the shell model has also been earlier connected to the shell model though a unitary transformation \cite{Castanos1992a,Castanos1992b,Castanos1994}, albeit a more involved one. 
The crucial role of the spin-orbit interaction in formulating the nuclear magic numbers and in paving the way for the proxy-SU(3) approximation has been discussed in detail in the article in which proxy-SU(3) has been introduced \cite{Bonatsos2017a}, as well as in relation to the connection of the proxy-SU(3) scheme to the shell model \cite{Martinou2020} and in the review article of Ref. \cite{Bonatsos2023}.

A mapping \cite{Castanos1988,Draayer1989} between the invariants of the collective model of Bohr and Mottelson and the invariants of SU(3) leads to expressions of the collective variables $\beta$ and $\gamma$ in terms of the Elliott quantum numbers $\lambda$ and $\mu$, characterizing the irreducible represenatiions $(\lambda,\mu)$ of SU(3) in the Elliott notation \cite{Elliott1958a,Harvey1968}. It turns out that in this way parameter-free predictions for the collective variables $\beta$ and $\gamma$ can be produced for all nuclei lying not too close to closed shells \cite{Bonatsos2017b}, leading to useful physical results, like the justification of the dominance of the prolate over oblate shapes in the ground states of even-even nuclei \cite{Bonatsos2017b,Bonatsos2017c}, and the transition from prolate to oblate shapes in the rare earth region around $N=114$ \cite{Bonatsos2017b,Bonatsos2024a}.  

In the present article we present full numerical results for all nuclei in the regions ($Z=28$-50, $N=28$-126) and ($Z=50$-82, $N=50$-184), covering all known nuclei in these regions and even going beyond the relevant proton and neutron driplines \cite{Wang2015,Neufcourt2020,Yang2021,Chai2022,Ahn2024}. The purpose of the present study is to offer a uniform set of predictions which can be used for many purposes, ranging from comparison with predictions by other models to drawing physical conclusions about the structural characteristics in various regions of the nuclear chart. Some fragmented results have already appeared in Ref. \cite{Bonatsos2024b}, summerized in Fig. 1 of Ref. \cite{Bonatsos2024b}.

In Section II the irreducible representations (irreps) of SU(3) needed for this calculation are constructed, while in Section III the numerical values of the collective variables $\beta$ and $\gamma$ for these irreps are tabulated. Some examples of use of the numerical results are given in Section IV, while in Section V the conclusions of the present work and the relevant outlook are discussed.  

\section{Irreducible representations} \label{hw} 

Proxy-SU(3) symmetry is based on the Pauli principle and the short-range nature of the nucleon-nucleon interaction, which pushes the nuclear system towards the most symmetric irreducible representation allowed by the Pauli principle \cite{Martinou2021b}, called the highest weight irreducible representation (hw irrep) in group theoretical language. While it turns out that the hw irreps alone lead to successful predictions for several physical quantities \cite{Bonatsos2017b,Bonatsos2023}, in some cases the next highest weight irreps (nhw irreps) should also be taken into account \cite{Bonatsos2024b}. 

The irreps appearing in the U(N)$\supset$SU(3) decomposition for $N=6$, 10, 15, and 21, corresponding to the proxy-SU(3) $sd$, $pf$, $sdg$, $pfh$, and $sdgi$  shells respectively, have been calculated using the code of Ref. \cite{Draayer1989a}, while a newer code also exists \cite{Langr2019} (see also \cite{Alex2011} for an alternative). A simple formula providing the hw irreps alone has been given by Kota \cite{Kota2018}, not only for identical nucleons, but also for protons and neutrons occupying the same major shell, having good spin-isospin SU(4) symmetry. Full SU(3) decompositions for nuclei having $32 \leq Z, N \leq 46$ 
with protons and neutrons occupying the same shell and possessing the proxy-SU(4) symmetry \cite{Kota2024} have been given in Ref. \cite{Kota2025}. 

In Tables I and II the results for U(6) and U(10) are shown. The irreps are listed in order of decreasing weight, thus the hw irrep is the first one, the next hw irrep is the second one, and so on. Only results for even number of particles $M$ are shown, since up to now the proxy-SU(3) approach has been used only for even-even nuclei. In addition, only the irreps with even $\lambda$ and even $\mu$ are shown, since we are interested in collective bands with $K=0$, 2, 4, \dots, in which only even $\lambda$ and even $\mu$ appear. In cases in which an irrep appears more than once in the given decomposition, its multiplicity is given as an exponent.  

In Tables III, IV, and V  the results for U(15), U(21), and U(28) are shown. Since the number of irreps appearing in each U($N$) decomposition increases rapidly with $N$, only the first 8 irreps appearing for each particle number $M$ are given in Tables III, IV, and V. 

The results appearing in Tables II and III can be compared to the results of the full decompositions listed in Tables I and II of Ref. \cite{Kota2025}, calculated through a different approach. In Ref. \cite{Kota2025} all irreps for all $\lambda$ and $\mu$, even or odd, are listed, for all particle numbers $M$, even or odd. 
The comparison shows that the irreps listed in the present Tables II and III are in agreement with the corresponding irreps in Tables I and II of Ref. \cite{Kota2025}, showing that the truncation made in the proxy-SU(3) framework does not affect the symmetry of the collective bands under discussion. 

In what follows, only the hw and nhw irreps will be employed, collected in Table VI for convenience. 

In order to determine the hw irrep for a given nucleus, one needs the hw irrep for its valence protons, $(\lambda_p,\mu_p)$, and the hw irrep for its valence neutrons,  $(\lambda_n,\mu_n)$. These are found from Table VI. Then the most stretched irrep, $(\lambda_p+\lambda_n, \mu_p+\mu_n)$ is the hw irrep characterizing the nucleus. There is no ambiguity in its determination, among other reasons because the hw irreps always have multiplicity equal to one. 

As an example, one may consider the nucleus \isotope[166][68]{Er}$_{98}$, which has $68-50=18$ valence protons in the 50-82 shell, corresponding to U(15) within the proxy-SU(3) scheme, which correspond to the hw irrep (18,6) in Table VI, as well as $98-82=16$ valence neutrons in the 82-126 shell, corresponding to U(21) within the proxy-SU(3) scheme, which correspond to the hw irrep (34,8) in Table VI. Therefore the hw irrep for \isotope[166][68]{Er}$_{98}$ turns out to be 
$(18+34, 6+8)=(52,14)$.

The determination of the nhw irrep for \isotope[166][68]{Er}$_{98}$ is only slightly longer. One has to consider the hw and nhw irreps for the 18 valence protons in U(15), which are $P_1$=(18,6) and $P_2$=(20,2) respectively, as well as the hw and nhw irreps for the 16 valence neutrons in U(21), which are $N_1$=(34,8) and $N_2$=(36,4) respectively. Obviously there are 4 possible combinations, $P_1+N_1$, which gives the hw irrep of \isotope[166][68]{Er}$_{98}$, as well as $P_1+N_2=(54,10)$,  $P_2+N_1=(54,10)$, and  $P_2+N_2=(56,6)$. The rule for selecting the irrep with the highest weight among them, is given in Ref. \cite{Hecht1965}. The highest weight irrep $(\lambda,\mu)$ is characterized by the highest value of $2\lambda+\mu$, while among irreps with equal values of $2\lambda+\mu$, the irrep with the highest $\mu$ wins. In the present case, all irreps have $2\lambda+\mu=118$, therefore (52,14) is the hw irrep, as expected, while (54,10) is the nhw irrep. Notice that the nhw irrep is given by $(\lambda+2,\mu-4)$ if the hw irrep is $(\lambda,\mu)$. 

The results for nuclei with ($Z=28$-50, $N=28$-126) and ($Z=50$-82, $N=50$-184) are given in Appendices A and B respectively. Several comments are in place. 

Looking at Appendix A, one sees that the hw irrep $(\lambda,\mu)$ is accompanied by the nhw irrep $(\lambda+2,\mu-4)$ for all U(N), except in the cases of the particle numbers $M=2$, 4, 6, 12, 20, 30. As a consequence, in nuclei in which both the valence protons and the valence neutrons avoid the numbers $M=2$, 4, 6, 12, 20, 30, the nhw irrep accompanying the hw irrep $(\lambda,\mu)$ will be $(\lambda+2,\mu-4)$. If either the valence protons or the valence neutrons numbers coincide with one of the numbers $M=2$, 4, 6, 12, 20, 30, the nhw irrep will not necessarily be following this rule, and has to be determined according to the above mentioned rules of Ref. \cite{Hecht1965}, by selecting the irrep with the highest value of $2\lambda+\mu$, and among irreps sharing the same value of $2\lambda+\mu$ the irrep with the highest value of $\mu$. 

It turns out that in the cases in which only one of the valence numbers (protons or neutrons) coincides with $M=2$, 4, 6, 12, 20, 30, the rules of Ref. \cite{Hecht1965} restore the above rule, namely that the nhw irrep is given by $(\lambda+2,\mu-4)$ if the hw irrep is $(\lambda,\mu)$. Only in the cases in which both the valence protons and the valence neutrons numbers coincide with any of the numbers  $M=2$, 4, 6, 12, 20, 30 is the rule broken, as we shall demonstrate through two examples. 

Let us first consider the nucleus \isotope[170][68]{Er}$_{102}$. The hw and nhw irreps for the 18 valence protons in U(15) are $P_1$=(18,6) and $P_2$=(20,2), as above.
However, the $102-82=20$ valence neutrons coincide with one of the numbers in the list of  $M=2$, 4, 6, 12, 20, 30, giving from the U(21) columns of Table VI 
$N_1=(40,0)$ and $N_2=(10,14)$. The 4 possible combinations are then, $P_1+N_1$, which gives the hw irrep (58,6) of the \isotope[170][68]{Er}$_{102}$, as well as $P_1+N_2=(48,20)$,  $P_2+N_1=(60,2)$, and  $P_2+N_2=(50,16)$. The rule for selecting the irrep with the highest weight among them \cite{Hecht1965} has been mentioned above. In the present case, the irreps (58,6) and (60,2) have $2\lambda+\mu=122$, while the irreps (48,20) and (50,16) have $2\lambda+\mu=116$.
Therefore the first two irreps have priority for being the hw irrep, since they have the highest value of $2\lambda+\mu$. Among them, (58,6) is the hw irrep, as expected, since it has higher $\mu$ than the irrep (60,2), while the irrep (60,2) is the nhw irrep. We remark that the rule of the hw irrep $(\lambda,\mu)$ followed by the nhw irrep $(\lambda+2,\mu-4)$ is restored. 

As a second example, let us consider the nucleus \isotope[164][70]{Yb}$_{94}$, for which both the $70-50=20$ valence protons and the $94-82=12$ valence neutrons belong to the $M=2$, 4, 6, 12, 20, 30 list. For the 20 valence protons in the 50-82 shell we see in the U(15) columns of Table VI that the hw and nhw irreps are $P_1=(20,0)$ and $P_2=(10,14)$, while for the 12 valence neutrons in the 82-126 shell we see in the U(21) columns of Table VI that the hw and nhw irreps are $N_1=(36,0)$ and $N_2=(28,10)$. The 4 possible combinations are then, $P_1+N_1$, which gives the hw irrep (56,0) of the \isotope[164][70]{Yb}$_{94}$, as well as $P_1+N_2=(48,10)$,  $P_2+N_1=(46,14)$, and  $P_2+N_2=(38,24)$. We remark that the quantity $2\lambda+\mu$ for these 4 irreps obtains the values 112, 106, 106, 100 respectively. Therefore (56,0), which possesses the highest $2\lambda+\mu$ value, is the hw irrep, according to the above mentioned rules of Ref. \cite{Hecht1965}, while the irreps (48,10) and (46,14), which have the same $2\lambda+\mu$ value, compete for the nhw irrep place, the winner, according to the rules of Ref. \cite{Hecht1965}, being (46,14), since it possesses the highest $\mu$ value between them. We see that the hw irrep (56,0) and the nhw irrep (46,14) do not follow the rule that the nhw is given by $(\lambda+2,\mu-4)$ if the hw irrep is $(\lambda,\mu)$.

For $M=2$, 6, 12, 20, 30, a mathematical proof exists that in any U(N) the hw irrep will have $\mu=0$. The proof is given in Appendix C. Therefore, their presence in the list of $M=2$, 4, 6, 12, 20, 30 is easily justified. The presence of $M=4$ in this list has a somewhat different origin. It is based on the fact that for any U(N), $M=4$ is the only number of identical particles for which hw irreps with $\mu=2$ occur, as it can be seen in Table VI. 
The consequences of the appearance of $\mu=2$ for either the valence protons or the valence neutrons of a nucleus can be clarified through two examples. 

As a first example, let us consider the nucleus \isotope[154][68]{Er}$_{86}$. The hw and nhw irreps for the 18 valence protons in U(15) are $P_1$=(18,6) and $P_2$=(20,2), as above. However, the $86-82=4$ valence neutrons give from the U(21) columns of Table VI $N_1=(16,2)$ and $N_2=(12,4)$. The 4 possible combinations are then, $P_1+N_1$, which gives the hw irrep (34,8) of the \isotope[154][68]{Er}$_{86}$, as well as $P_1+N_2=(30,10)$,  $P_2+N_1=(36,4)$, and  $P_2+N_2=(32,6)$. The rule for selecting the irrep with the highest weight among them \cite{Hecht1965} has been mentioned above. In the present case, the irreps (34,8) and (36,4) have $2\lambda+\mu=76$, while the irreps (30,10) and (32,6) have $2\lambda+\mu=70$. Therefore the first two irreps have priority for being the hw irrep, since they possess the highest $2\lambda+\mu$ value. Then (34,8) is the hw irrep, as expected, since it has higher $\mu$ than the irrep (36,4), while the irrep (36,4) is the nhw irrep. We remark that the rule of the hw irrep $(\lambda,\mu)$ followed by the nhw irrep $(\lambda+2,\mu-4)$ is restored. 

As a second example, let us consider the nucleus \isotope[156][70]{Yb}$_{86}$, for which both the $70-50=20$ valence protons and the $86-82=4$ valence neutrons belong to the $M=2$, 4, 6, 12, 20, 30 list. For the 20 valence protons in the 50-82 shell we see in the U(15) columns of Table VI that the hw and nhw irreps are $P_1=(20,0)$ and $P_2=(10,14)$, while for the 4 valence neutrons in the 82-126 shell we see in the U(21) columns of Table VI that the hw and nhw irreps are $N_1=(16,2)$ and $N_2=(12,4)$. The 4 possible combinations are then, $P_1+N_1$, which gives the hw irrep (36,2) of the \isotope[156][70]{Yb}$_{86}$, as well as $P_1+N_2=(32,4)$,  $P_2+N_1=(26,16)$, and  $P_2+N_2=(22,18)$. We remark that the quantity $2\lambda+\mu$ for these 4 irreps obtains the values 74, 68, 68, 62 respectively. Therefore (36,2), which possesses the highest $2\lambda+\mu$ value, is the hw irrep, according to the above mentioned rules of Ref. \cite{Hecht1965}, while the irreps (32,4) and (26,16), which have the same $2\lambda+\mu$ value, compete for the nhw irrep place, the winner, according to the rules of Ref. \cite{Hecht1965}, being (26,16), since it possesses the highest $\mu$ value between them. We see that the hw irrep (36,2) and the nhw irrep (26,16) do not follow the rule that the nhw irrep is given by $(\lambda+2,\mu-4)$ if the hw irrep is $(\lambda,\mu)$.

We see that the occurrence of $\mu=2$ in the hw irrep of $M=4$, has exactly the same consequences as the occurrence of $\mu=0$ in the hw irrep of $M=2$, 6, 12, 20, 30, as far as the rule that the nhw irrep is given by $(\lambda+2,\mu-4)$ if the hw irrep is $(\lambda,\mu)$ is concerned. 

In general, in nuclei in which both the valence protons and the valence neutrons belong to the list $M=2$, 4, 6, 12, 20, 30, the nhw does not follow the rule that the nhw is given by $(\lambda+2,\mu-4)$ if the hw irrep is $(\lambda,\mu)$. From the mathematical point of view this is expected, since the hw irrep turns out to have $\mu=0$, therefore $\mu-4$ cannot occur. From the physics point of view, in these cases the hw irrep can accommodate only the ground state band, which is unphysical, since experimentally the ground state band (the lowest $K=0$ band) and the $\gamma_1$ band (the lowest $K=2$ band) are expected to belong to the same SU(3) irrep, since they are connected by strong interband B(E2) transition rates \cite{ensdf,Bonatsos2025} and bear many structural similarities \cite{Jolos2006,Jolos2007,Minkov1997,Minkov1999,Minkov2000,Bonatsos2021}. Therefore it becomes clear that some mixing of the hw irrep and the nhw irrep will become necessary in such cases \cite{Bonatsos2024b}. 

The last point demonstrates the basic difference between the predictions of the SU(3) symmetry of the IBM and the proxy-SU(3) symmetry for deformed nuclei.
In the IBM \cite{Arima1978,Iachello1987}, the lowest lying irrep is $(2N,0)$, where $N$ is the number of bosons, coming from the pairs of valence protons and valence neutrons, each counted from the nearest closed shell, while the next lowest lying irrep is $(2N-4,2)$. As a consequence, the lowest lying irrep can accommodate only the ground state band (gsb), which has $K=0$, while the quasi-$\gamma_1$ band (lowest $K=2$ band) and the
quasi-$\beta_1$ band (second lowest $K=0$ band) will be accommodated in the $(2N-4,2)$ irrep. Given the  fact that the experimental interband B(E2) transition rates connecting the $\gamma_1$ band to the gsb are relatively strong \cite{ensdf,Bonatsos2025} and that the gsb and the 
$\gamma_1$ band have several structural similarities \cite{Jolos2006,Jolos2007,Minkov1999,Minkov2000,Bonatsos2021}, one has to break the SU(3) symmetry in order to accommodate them within the IBM scheme, since B(E2) transitions between different irreps are forbidden. This breaking of the SU(3) symmetry is not necessary within the proxy-SU(3) scheme, since the quasi-$\gamma_1$ band and the gsb belong to the same irrep, the hw irrep, within which interband transitions are allowed. The difference is rooted in the fact that IBM uses bosons, while proxy-SU(3) uses fermions. This difference has been earlier pointed out in the framework of the pseudo-SU(3) symmetry approximation to the shell model \cite{Draayer1982,Draayer1983,Draayer1984,Draayer1993} which is also formulated in terms of fermions.    

\section{Collective deformation parameters}

The collective deformation variables $\beta$ and $\gamma$ can be obtained as functions of the Elliott quantum numbers $\lambda$ and $\mu$ 
through the established mapping \cite{Castanos1988,Draayer1989} between the invariants of the Bohr collective model \cite{Bohr1998b} 
and the invariants of SU(3) \cite{Iachello2006}, which are its Casimir operators of second and third order, $C_2$ and $C_3$.     
According to this mapping the $\gamma$ variable is given by \cite{Castanos1988,Draayer1989}
\begin{equation}\label{g1}
\gamma = \arctan \left( {\sqrt{3} (\mu+1) \over 2\lambda+\mu+3}  \right),
\end{equation}
while the $\beta$ variable is related to the second order Casimir operator of SU(3), the eigenvalues of which are 
\cite{Kota2020}
  \begin{equation}\label{C2} 
 C_2(\lambda,\mu)= (\lambda^2+\lambda \mu + \mu^2+ 3\lambda +3 \mu), 
\end{equation}
and is given by \cite{Castanos1988,Draayer1989}
\begin{equation}\label{b1}
	\beta^2= {4\pi \over 5} {1\over (A \bar{r^2})^2} (\lambda^2+\lambda \mu + \mu^2+ 3\lambda +3 \mu +3), 
\end{equation}
where $A$ is the mass number of the nucleus, while $\bar{r^2}$ is related to the dimensionless mean square radius \cite{Ring1980}, $\sqrt{\bar{r^2}}= r_0 A^{1/6}$. The dimensionless mean square radius is obtained by dividing the mean square radius, which grows as $A^{1/3}$, by the oscillator length, which is proportional to $A^{1/6}$ \cite{Ring1980}. The constant $r_0$ is found from a fit over a wide range of nuclei \cite{DeVries1987,Stone2014} to have the value $r_0=0.87$. 

We remark that $\beta^2$ is proportional to $C_2+3$. Taking into account that only the valence shells have been considered, the values of $\beta$ should be multiplied by a scaling factor $A/(S_p+S_n)$, where $S_p$ ($S_n$) is the size of the proton (neutron) valence shell \cite{Bonatsos2017b}. For example, in the case of the rare earth region, in which the valence protons lie in the 50-82 shell and the valence neutrons lie in the 82-126 shell, one has $S_p=32$ and $S_n=44$, thus the scaling factor is $A/76$.  

The deformation parameters corresponding to the hw and nhw irreps for nuclei with ($Z=28$-50, $N=28$-126) and ($Z=50$-82, $N=50$-184) are given in Appendices A and B  respectively. Several comments are in place. 

As explained in the previous section, for most nuclei the hw irrep $(\lambda,\mu)$ is accompanied by the nhw irrep $(\lambda+2,\mu-4)$. Elementary calculations show that in Eq. (\ref{b1}), if the Casimir operator for the hw irrep, $C_{hw}$, is given by Eq. (\ref{C2}), then the Casimir operator of the nhw irrep, 
$C_{nhw}$, is given by $C_{nhw}=C_{hw} -6(\mu-1)$. Since in most cases $C_{hw}$ is much larger than $\mu$, the change caused in $\beta$ when passing from the hw irrep to the nhw irrep is relatively small. This result supports the robustness of the proxy-SU(3) approximation, justifying why its parameter-independent predictions given by the hw irrep suffice to provide sufficient agreement to the empirical values of $\beta$, as already seen in Refs. \cite{Bonatsos2017b,Bonatsos2023}. The Pauli principle, combined with the short-range nature of the nucleon-nucleon interaction, pushes the nucleus to the hw irrep, which provides a certain prediction for $\beta$. Even if mixing with the nhw irrep were required, the prediction for $\beta$ would have been very little affected. 

A rather similar situation occurs for the collective variable $\gamma$. As one can see in Eq. (\ref{g1}), in the cases in which the hw irrep $(\lambda,\mu)$ is accompanied by the nhw irrep $(\lambda+2,\mu-4)$, the denominator in Eq. (\ref{g1}) remains the same, while in the numerator the factor $(\mu+1)$ of the hw case becomes $(\mu-3)$ in the nhw case. Given the fact that the denominator in Eq. (\ref{g1}) is usually much larger than $\mu$, the change inflicted is in most cases rather small, again supporting the robustness of the proxy-SU(3) predictions. 

The previous two paragraphs indicate that in many nuclei the nhw irrep has a somewhat lower $\beta$ and a somewhat lower $\gamma$ than the hw irrep. In other words the nhw is a little less deformed and a little more prolate than the hw. As a consequence, the second $K=0$ band (the 
quasi-$\beta_1$) band, is expected to be   a little less deformed and a little more prolate than the gsb. This fact clearly shows that shape coexistence \cite{Heyde1983,Wood1992,Heyde2011,Heyde2016,Garrett2022,Bonatsos2023a} cannot be related to the nhw irrep. The collective band coexisting with the gsb should have a radically different shape, indicating that it has to be attributed to a different mechanism. Indeed, a dual-shell mechanism \cite{Martinou2021,Martinou2023,Bonatsos2023b} has been developed for the description of shape coexistence within the proxy-SU(3) scheme, supported by both covariant  density functional theory \cite{Bonatsos2022a,Bonatsos2022b} and non-relativistic mean-field \cite{Hasan2026} calculations. 

The similarity between the gsb and the first excited $K=0$ band (the quasi-$\beta_1$ band) could also be useful for determining the collective $0_2^+$ state among the many low-lying $0^+$ states seen experimentally in several deformed nuclei 
\cite{Asai1997,Aprahamian2002,Lesher2002,Wirth2004,Bucurescu2006,Meyer2006a,Meyer2006b,Suliman2008,Lesher2007,Balabanski2011,Aprahamian2017}. The nature of the $0_2^+$ states has been an open problem for a long time \cite{Garrett2001,Aprahamian2025}. While in deformed nuclei the $2_1^+$ state belongs to the gsb and the $2_2^+$ state appears to be the bandhead of the collective first $K=2$ band (the quasi-$\gamma_1$ band), connected to the gsb by relatively strong B(E2) transition rates \cite{ensdf,Bonatsos2024b}, no such general rule exists for the first excited $0^+$ state, $0_2^+$. Furthermore, the vibrational collective character of this state has been questioned during the last several years \cite{SharpeySchafer2008,SharpeySchafer2010,SharpeySchafer2011a,SharpeySchafer2011b,SharpeySchafer2019}. It might therefore 
be interesting to examine to what extent the nature of the $0_2^+$ states predicted by the proxy-SU(3) symmetry appears experimentally. 

From Eq. (\ref{g1}) it is evident that $\mu=0$ leads to very low values of $\gamma$. As a consequence, when plotting the values of $\gamma$ predicted by 
proxy-SU(3) for a series of isotopes or isotones, one sees deep minima at the valence numbers  $M=2$, 4, 6, 12, 20, 30 discussed above (see, for example, Fig. 5 of Ref. \cite{Bonatsos2017b}), which are not seen in the empirical values of $\gamma$ extracted from the data (see, for example, section VI of Ref. \cite{Bonatsos2017b} for details). Agreement to the empirical values can be restored by mixing the hw irrep with the nhw irrep (see, for example, Fig. 2 of Ref. \cite{Bonatsos2024b}). This restoration becomes possible because of the rules of Ref. \cite{Hecht1965}. The rule that, among the irreps possessing the same value of $2\lambda+\mu$, the one with the highest $\mu$ becomes the nhw irrep, guarantees that the nhw irrep accompanying a hw irrep with $\mu=0$ will be characterized by a much higher value of $\mu$, thus substantially raising the value of $\gamma$ for the relevant nucleus. 

As an example, consider the above mentioned nucleus \isotope[164][70]{Yb}$_{94}$. The hw irrep (56,0) yields $\gamma_{hw}=0.86^{\rm o}$, while the nhw irrep (46,14) gives $\gamma_{nhw}=13.41^{\rm o}$. Assuming, as a crude approximation, mixing of the two irreps with 50\% participation of each of them, the average value of $\gamma$ becomes 7.14$^{\rm o}$, which is of the same order as the $\gamma_{hw}$ values of the neighboring nuclei with $N=90$, 92, 96, 98, which are 5.00$^{\rm o}$, 4.63$^{\rm o}$, 5.92$^{\rm o}$,  and 7.46$^{\rm o}$ respectively.   

It should be mentioned at this point that the calculated values of $\beta$ and $\gamma$ are parameter-free in the sense that no adjustable fit parameters appear in their calculation. However, they do depend on the assumptions and approximations made in the formulation of the proxy-SU(3) scheme, 
reviewed in Ref. \cite{Bonatsos2023}, as well as on the scaling factor involved in Eq. (\ref{b1}) mentioned earlier.

It should also be mentioned that calculations for the collective variables $\beta$ and $\gamma$ within the pseudo-SU(3) framework \cite{Arima1969,Hecht1969,RatnaRaju1973,Draayer1982,Draayer1983,Draayer1984,Bahri1992,Ginocchio1997} provide numerical results \cite{Bonatsos2020a} qualitatively similar to the proxy-SU(3) results, provided that the relevant hw irreps are used in both cases. 
It is quite encouraging to see that two different approximations, based on different assumptions and different unitary transformations, using different sets of valence protons and neutrons, and therefore ending up with very different hw irreps for the same nucleus, provide similar values for the collective variables, indicating that the SU(3) symmetry, the Pauli principle, and the short-range nature of the nucleon-nucleon interaction are the important factors shaping up the collective nuclear properties, irrespectively of technical details.  

\section{Examples}

\subsection{Triaxiality in $^{104}$Ru}

In the Interacting Boson Model-2 (IBM-2) \cite{Iachello1987,Talmi1993,Frank2005}, triaxiality in medium-mass nuclei is assumed to occur when the valence protons correspond to holes (thus corresponding to oblate irreps of the form $(0,2N_\pi)$, where $N_\pi$ stands for the number of  pairs of proton holes, counted from the nearest closed shell, while the valence neutrons correspond to particles (thus corresponding to prolate irreps of the form $(2N_\nu, 0)$, where $N_\pi$ stands for the number of pairs of neutron particles, counted from the nearest closed shell \cite{Dieperink1982,Dieperink1983,Dieperink1984,Iachello1984,Dieperink1985,Walet1987}. The total SU(3) irrep is then $(2N_\nu, 2N_\pi)$. In \isotope[104][44]{Ru}$_{60}$, which is the textbook example of this case \cite{Dieperink1982}, there are 6 valence proton holes, corresponding to $N_\pi=3$, and 10 neutron particles, corresponding to $N_\nu=10$.  Therefore \isotope[104][44]{Ru}$_{60}$ is represented within IBM-2 by the irrep (10,6), which, through Eq. (\ref{g1}) gives $\gamma=22.7^{\rm o}$. 

Within proxy-SU(3), in which valence particles are always counted from the closed shell below, there are 16 valence protons within the 28-50 shell, which is characterized by the U(10) symmetry, thus from Table VI they correspond to the (2,8) irrep, as well as 10 valence neutrons in the 50-82 shell, which is characterized by the U(21) symmetry, thus from Table VI they correspond to the (20,4) irrep. Therefore the total proxy-SU(3) irrep for \isotope[104][44]{Ru}$_{60}$ is (22,12), which, as seen in Appendix A, corresponds to $\gamma=20.9^{\rm o}$. The near agreement between the predictions for $\gamma$ by the two radically different approaches, IBM-2 and proxy-SU(3), which employ different assumptions and approximations, is remarkable.   

\subsection{Evolution of collective variables along the valley of stability} \label{valley} 

The valley of stability, recently used as a test-ground for a potential-energy-surface (PES) approach \cite{Meng2022}, can be used for testing the proxy-SU(3) predictions as well. 

The nuclei along the stability line are listed in Table VII, obtained through the Green's formula \cite{Green1955}
\begin{equation}
N-Z = 0.4 {A^2 \over A+200}.  
\end{equation}  

In Fig. 1(a) the proxy-SU(3) predictions for the collective variable $\beta$ along the valley of stability, obtained with the hw irrep,  are compared to the empirical values obtained from experimental values of the transition rates $B(E2; 0_1^+\to 2_1^+)$ \cite{Pritychenko2016}, with quite good agreement seen. 

In Fig. 1(b)  the proxy-SU(3) predictions for the collective variable $\gamma$ along the valley of stability, obtained with the hw irrep, are given. We see that deep minima occur at the proton numbers 32, 40, 52, 62, 70, which represent 4 and 12 protons above the magic number 28, as well as 2, 12, 20 protons above the magic number 50, corresponding to the collection of 2, 4, 6, 12, 20, 30 valence protons, related to irreps with $\mu=0$, as seen in Sec. \ref{hw}. 

In Fig. 1(c) the proxy-SU(3) predictions for the collective variable $\gamma$ at the deep minima, which are unphysical as discussed in Sec. \ref{hw}, have been replaced by the average value of $\gamma$ obtained from the hw irrep and the nhw irrep, as discussed in Sec. \ref{hw}, resulting in a smoother curve.  

In addition in Figs. 1(b) and 1(c) empirical values of $\gamma$ for the nuclei in which they are available are shown. A way to extract $\gamma$ values 
from the data is based on the Davydov model \cite{Davydov1958,Davydov1959}, using the energy ratio of the bandhead of the quasi-$\gamma_1$ band, $2_2^+$, over the first excited state of the ground state band, $2_1^+$ 
\begin{equation} \label{R}
R= {E(2_2^+) \over E(2_1^+)} 
\end{equation}
through the expression \cite{Casten2000}
\begin{equation} \label{g}
\gamma = {1\over 3} \sin^{-1} \left( {3\over R+1} \sqrt{R\over 2}  \right). 
\end{equation}
This formula works for $R\geq 2$. Furthermore, since it regards a triaxial rotator, it is expected, as already remarked in Ref. \cite{Varshni1970}, to be applicable for nuclei with ratio 
\begin{equation} \label{R42}
R_{4/2}={E(4_1^+) \over E(2_1^+)} 
\end{equation}
above the 8/3=2.667 value, which corresponds to the rigid triaxial rotator, up to 10/3=3.333, which corresponds to the rigid axial rotator \cite{Casten2000}. 

Nuclei with experimentally known ratios $R>2$ and $R_{4/2}>8/3$, taken from the ENSDF database \cite{ensdf},  are shown in Table VIII, along with the $\gamma_R$ values produced through Eq. (\ref{g}). 
In addition, the empirical values $\gamma_{TR}$, obtained recently though a method involving two $E2$  matrix elements \cite{Lawrie2025}, as well as the empirical values $\gamma_{KC}$, obtained \cite{Lawrie2025} through the Kumar-Cline method \cite{Kumar1972,Cline1986} are included for comparison.
In Figs. 1(b) and 1(c) we see that the empirical values are in qualitative agreement among themselves, as well as with the proxy-SU(3) predictions. 
The need for taking into account the nhw irrep in addition to the hw irrep at $Z=70$ is clearly supported by the empirical values, as seen from the comparison between Figs. 1(b) and 1(c).     

\subsection{Mirror symmetry} \label{mirror}

The valence mirror symmetry between the $Z=50$ isotopes and the $N=82$ isotones has been recently studied \cite{Zong2024} in the framework of the nucleon-pair approximation (NPA) \cite{Zhao2014}. It would be of interest to see to what extent such a mirror symmetry occurs just above the $Z=50$ and $N=82$ magic numbers. 

In Fig. 2(a) the proxy-SU(3) predictions for the collective variable $\beta$ are shown for $Z=52$, 54, 56 and $N=52$-80, as well as for $N=84$, 86, 88 and $Z=52$-80. In other words, we consider up to three valence proton pairs above the $Z=50$ shell closure, as well as  up to three valence neutron pairs above the $N=82$ shell closure. Very close agreement is seen between the $Z=52$ and $N=86$ predictions, as well as between the $Z=54$ and $N=88$ predictions, especially around the middle of the 50-82 interval, where maximum quadrupole deformation occurs.

We see that a certain degree of similarity exists between the nuclei with two (four) protons outside the $Z=50$ shell and the nuclei with four (six) neutrons outside the $N=82$ shell. It is of interest to examine if this similarity also appears for the corresponding holes, i.e., for nuclei with
two (four) protons holes below the $Z=82$ shell and nuclei with four (six) neutrons holes below the $N=126$ shell.
The relevant plot is seen in Fig. 2(b). We see that indeed similarity is seen between the $Z=80$ and $N=122$ predictions, as well as between the 
$Z=78$ and $N=120$ predictions, especially below $M=70$ particles. However, the $\beta$ values occurring in the holes cases in Fig. 2(b) are lower than the ones occurring in the corresponding particles cases in Fig. 2(a). 

We are now going to examine to what extent such a mirror symmetry occurs just above the $Z=28$ and $N=50$ magic numbers. 

In Fig. 3(a) the proxy-SU(3) predictions for the collective variable $\beta$ are shown for $Z=30$, 32, 34 and $N=30$-48, as well as for $N=52$, 54, 56 and $Z=30$-48. i.e.,  up to three valence proton pairs above the $Z=28$ shell closure, as well as  up to three valence neutron pairs above the $N=50$ shell closure are considered. Very close agreement is seen between the $Z=30$ and $N=54$ predictions, as well as between the $Z=32$ and $N=56$ predictions, especially around the middle of the 28-50 interval, where maximum quadrupole deformation occurs.

We see again that similarity is seen between the nuclei with two (four) protons outside the $Z=28$ shell and the nuclei with four (six) neutrons outside the $N=50$ shell. Therefore we are going to examine also in the present case if this similarity also appears for the corresponding holes, i.e., for nuclei with two (four) protons holes below the $Z=50$ shell and nuclei with four (six) neutrons holes below the $N=82$ shell.
The relevant plot is seen in Fig. 3(b). We see that indeed similarity is seen between the $Z=48$ and $N=78$ predictions, as well as between the 
$Z=46$ and $N=76$ predictions, especially below $M=40$ particles. In analogy to the heavier shells considered above, the $\beta$ values occurring in the holes cases in Fig. 3(b) are lower than the ones occurring in the corresponding particles cases in Fig. 3(a).

In conclusion, we see in both cases that a certain degree of similarity exists between the nuclei with two (four) protons outside the proton closed  shell and the nuclei with four (six) neutrons outside the neutron closed shell. The same picture appears between the nuclei with two (four) proton holes inside the proton closed shell and the nuclei with four (six) neutron holes inside the neutron closed shell, although in the case of holes the predicted values of $\beta$ are systematically lower than the corresponding values for particles. In addition, the similarities are stronger below the 3D-HO magic numbers 40, 70 than above them. Further investigation is called for in order to locate the microscopic roots of these two observations.  

\section{Conclusions and outlook}

The hw SU(3) irreps and nhw SU(3) irreps in the framework of the proxy-SU(3) approximation to the shell model have been determined for all nuclei in the region 
of $Z=28-82$, $N=28-126$, and the corresponding parameter-free predictions for the $\beta$ and $\gamma$ collective variables are given. The numerical results have been used for making a connection to the IBM-2 in relation to the appearance of triaxiality in $^{104}$Ru, for examining the evolution of the collective variables along the valley of stability, as well as for testing mirror symmetries in medium-mass nuclei and rare earths.

It is hoped that the present tabulation of uniform proxy-SU(3) predictions in medium-mass and rare earth nuclei will facilitate comparisons to experimental data, as well as to the results of alternative theoretical approaches. Furthermore, a similar tabulation regarding actinide, superheavy and hyperheavy nuclei is in progress \cite{Bonatsos2026b}.   

Nuclei in which the valence protons and the valence neutrons belong to the same shell call for taking into account the isospin degree of freedom, giving rise to the proxy-SU(4) \cite{Kota2024} and pseudo-SU(4) \cite{VanIsacker2023} symmetries, taking advantage of the Wigner SU(4) symmetry \cite{Wigner1937}. As we have already seen, the hw irreps and nhw irreps for even-even nuclei within the approximations of the present work, coincide with the corresponding results obtained within the proxy-SU(4) symmetry. However, the present calculations are limited to bands with even $K$, for which only even values of $\lambda$ and $\mu$ are in use. The extension of the present approach to bands with odd $K$ is therefore desired, especially in the case of $N=Z$ nuclei, in which the existence of  $T=0$ proton-neutron pairs is currently attracting  much attention \cite{Liang2025,Palkanoglou2025}.

The extension of the proxy-SU(3) approach to the study of excitation spectra would be highly desirable. Preliminary work in this direction \cite{Bonatsos2019} indicates that the use of higher-order symmetry-preserving terms, like the three-body operator $\Omega$ and/or the four-body operator $\Lambda$ \cite{VandenBerghe1985,Vanthournout1990} (their mathematical names being $O_l^0$ and $Q_l^0$ respectively \cite{Hughes1973a,Hughes1973b,Judd1974,Partensky1979,DeMeyer1981a,DeMeyer1981b,VanderJeugt1983,DeMeyer1985,VanderJeugt1986,VanderJeugt1987}), would be needed in order to break the degeneracy between the levels of the ground state band and the first $K=2$ band (the quasi-$\gamma$ band), which, within the proxy-SU(3) scheme, belong to the same irrep, the hw one.

\section*{Acknowledgements} 

This research project is implemented in the framework of the Hellenic Foundation for Research and Innovation (H.F.R.I.) call ``3rd Call for H.F.R.I.'s Research Projects to Support Faculty Members and Researchers'' (H.F.R.I. Project Number: 23357). Support by the Bulgarian National Science Fund (BNSF) under Contract No. KP-06-N98/2 is gratefully acknowledged.

\section*{Appendix A}

In this Appendix the full tables of the highest weight (hw) irreducible representations of SU(3) and the next highest weight (nhw) irreps of SU(3) within the proxy-SU(3) scheme for nuclei in the $Z=28$-50, $N=28$-126 region are given, in order of increasing $Z$. The Elliott \cite{Elliott1958a} notation $(\lambda,\mu)$ is used for the SU(3) irreps. Proxy-SU(3) parameter-independent predictions for the collective variables $\beta$ and $\gamma$ within each irrep, calculated from Eqs. (\ref{b1}) and (\ref{g1}) are also given, labeled by hw and nhw respectively.

\begin{table}

\begin{tabular}{ r   r  r r c c r  }
\hline
nucleus  & hw & nhw & $\beta_{hw}$ & $\beta_{nhw}$ & $\gamma_{hw}$ & $\gamma_{nhw}$  \\ 
        
\hline

\isotope[60][30]{Zn}$_{30}$ & 12,  0 &  8,  2 & 0.165 & 0.132 &  3.67 & 13.90 \\
\isotope[62][30]{Zn}$_{32}$ & 14,  2 & 10,  4 & 0.201 & 0.171 &  8.95 & 17.78 \\
\isotope[64][30]{Zn}$_{34}$ & 18,  0 & 12,  6 & 0.232 & 0.209 &  2.54 & 20.17 \\
\isotope[66][30]{Zn}$_{36}$ & 16,  4 & 18,  0 & 0.235 & 0.230 & 12.52 &  2.54 \\
\isotope[68][30]{Zn}$_{38}$ & 16,  4 & 18,  0 & 0.233 & 0.228 & 12.52 &  2.54 \\
\isotope[70][30]{Zn}$_{40}$ & 18,  0 & 10, 10 & 0.226 & 0.220 &  2.54 & 30.00 \\
\isotope[72][30]{Zn}$_{42}$ & 12,  6 & 14,  2 & 0.201 & 0.191 & 20.17 &  8.95 \\
\hline
\end{tabular}

\end{table}

\begin{table}

\begin{tabular}{ r   r  r r c c r  }
\hline
nucleus  & hw & nhw & $\beta_{hw}$ & $\beta_{nhw}$ & $\gamma_{hw}$ & $\gamma_{nhw}$  \\ 
        
\hline  
                             
\isotope[62][32]{Ge}$_{30}$ & 14,  2 & 10,  4 & 0.201 & 0.171 &  8.95 & 17.78 \\
\isotope[64][32]{Ge}$_{32}$ & 16,  4 & 12,  6 & 0.238 & 0.209 & 12.52 & 20.17 \\
\isotope[66][32]{Ge}$_{34}$ & 20,  2 & 14,  8 & 0.267 & 0.247 &  6.59 & 21.79 \\
\isotope[68][32]{Ge}$_{36}$ & 18,  6 & 20,  2 & 0.272 & 0.264 & 15.08 &  6.59 \\
\isotope[70][32]{Ge}$_{38}$ & 18,  6 & 20,  2 & 0.269 & 0.262 & 15.08 &  6.59 \\
\isotope[72][32]{Ge}$_{40}$ & 20,  2 & 12, 12 & 0.259 & 0.258 &  6.59 & 30.00 \\
\isotope[74][32]{Ge}$_{42}$ & 14,  8 & 16,  4 & 0.238 & 0.227 & 21.79 & 12.52 \\
\hline
\end{tabular}

\end{table}

\begin{table}

\begin{tabular}{ r   r  r r c c r  }
\hline
nucleus  & hw & nhw & $\beta_{hw}$ & $\beta_{nhw}$ & $\gamma_{hw}$ & $\gamma_{nhw}$  \\ 
        
\hline  

\isotope[64][34]{Se}$_{30}$ & 18,  0 & 12,  6 & 0.232 & 0.209 &  2.54 & 20.17 \\
\isotope[66][34]{Se}$_{32}$ & 20,  2 & 14,  8 & 0.267 & 0.247 &  6.59 & 21.79 \\
\isotope[68][34]{Se}$_{34}$ & 24,  0 & 18,  6 & 0.298 & 0.272 &  1.95 & 15.08 \\
\isotope[70][34]{Se}$_{36}$ & 22,  4 & 24,  0 & 0.299 & 0.295 &  9.64 &  1.95 \\
\isotope[72][34]{Se}$_{38}$ & 22,  4 & 24,  0 & 0.296 & 0.292 &  9.64 &  1.95 \\
\isotope[74][34]{Se}$_{40}$ & 24,  0 & 16, 10 & 0.289 & 0.277 &  1.95 & 22.95 \\
\isotope[76][34]{Se}$_{42}$ & 18,  6 & 20,  2 & 0.262 & 0.255 & 15.08 &  6.59 \\
\hline
\end{tabular}

\end{table}

\begin{table}

\begin{tabular}{ r   r  r r c c r  }
\hline
nucleus  & hw & nhw & $\beta_{hw}$ & $\beta_{nhw}$ & $\gamma_{hw}$ & $\gamma_{nhw}$  \\ 
        
\hline

\isotope[66][36]{Kr}$_{30}$ & 16,  4 & 18,  0 & 0.235 & 0.230 & 12.52 &  2.54 \\
\isotope[68][36]{Kr}$_{32}$ & 18,  6 & 20,  2 & 0.272 & 0.264 & 15.08 &  6.59 \\
\isotope[70][36]{Kr}$_{34}$ & 22,  4 & 24,  0 & 0.299 & 0.295 &  9.64 &  1.95 \\
\isotope[72][36]{Kr}$_{36}$ & 20,  8 & 22,  4 & 0.305 & 0.296 & 17.00 &  9.64 \\
\isotope[74][36]{Kr}$_{38}$ & 20,  8 & 22,  4 & 0.302 & 0.293 & 17.00 &  9.64 \\
\isotope[76][36]{Kr}$_{40}$ & 22,  4 & 24,  0 & 0.291 & 0.287 &  9.64 &  1.95 \\
\isotope[78][36]{Kr}$_{42}$ & 16, 10 & 18,  6 & 0.272 & 0.260 & 22.95 & 15.08 \\
\hline
\end{tabular}

\end{table}

\begin{table}

\begin{tabular}{ r   r  r r c c r  }
\hline
nucleus  & hw & nhw & $\beta_{hw}$ & $\beta_{nhw}$ & $\gamma_{hw}$ & $\gamma_{nhw}$  \\ 
        
\hline

\isotope[68][38]{Sr}$_{30}$ & 16,  4 & 18,  0 & 0.233 & 0.228 & 12.52 &  2.54 \\
\isotope[70][38]{Sr}$_{32}$ & 18,  6 & 20,  2 & 0.269 & 0.262 & 15.08 &  6.59 \\
\isotope[72][38]{Sr}$_{34}$ & 22,  4 & 24,  0 & 0.296 & 0.292 &  9.64 &  1.95 \\
\isotope[74][38]{Sr}$_{36}$ & 20,  8 & 22,  4 & 0.302 & 0.293 & 17.00 &  9.64 \\
\isotope[76][38]{Sr}$_{38}$ & 20,  8 & 22,  4 & 0.300 & 0.291 & 17.00 &  9.64 \\
\isotope[78][38]{Sr}$_{40}$ & 22,  4 & 24,  0 & 0.288 & 0.284 &  9.64 &  1.95 \\
\isotope[80][38]{Sr}$_{42}$ & 16, 10 & 18,  6 & 0.270 & 0.257 & 22.95 & 15.08 \\
\hline
\end{tabular}

\end{table}

\begin{table}

\begin{tabular}{ r   r  r r c c r  }
\hline
nucleus  & hw & nhw & $\beta_{hw}$ & $\beta_{nhw}$ & $\gamma_{hw}$ & $\gamma_{nhw}$  \\ 
        
\hline

\isotope[70][40]{Zr}$_{30}$ & 18,  0 & 10, 10 & 0.226 & 0.220 &  2.54 & 30.00 \\
\isotope[72][40]{Zr}$_{32}$ & 20,  2 & 12, 12 & 0.259 & 0.258 &  6.59 & 30.00 \\
\isotope[74][40]{Zr}$_{34}$ & 24,  0 & 16, 10 & 0.289 & 0.277 &  1.95 & 22.95 \\
\isotope[76][40]{Zr}$_{36}$ & 22,  4 & 24,  0 & 0.291 & 0.287 &  9.64 &  1.95 \\
\isotope[78][40]{Zr}$_{38}$ & 22,  4 & 24,  0 & 0.288 & 0.284 &  9.64 &  1.95 \\
\isotope[80][40]{Zr}$_{40}$ & 24,  0 & 16, 10 & 0.282 & 0.270 &  1.95 & 22.95 \\
\isotope[82][40]{Zr}$_{42}$ & 18,  6 & 20,  2 & 0.255 & 0.248 & 15.08 &  6.59 \\
\hline
\end{tabular}

\end{table}

\setcounter{table}{5}

\begin{table}

\begin{tabular}{ r   r  r r c c r  }
\hline
nucleus  & hw & nhw & $\beta_{hw}$ & $\beta_{nhw}$ & $\gamma_{hw}$ & $\gamma_{nhw}$  \\ 
        
\hline

\isotope[72][42]{Mo}$_{30}$ & 12,  6 & 14,  2 & 0.201 & 0.191 & 20.17 &  8.95 \\
\isotope[74][42]{Mo}$_{32}$ & 14,  8 & 16,  4 & 0.238 & 0.227 & 21.79 & 12.52 \\
\isotope[76][42]{Mo}$_{34}$ & 18,  6 & 20,  2 & 0.262 & 0.255 & 15.08 &  6.59 \\
\isotope[78][42]{Mo}$_{36}$ & 16, 10 & 18,  6 & 0.272 & 0.260 & 22.95 & 15.08 \\
\isotope[80][42]{Mo}$_{38}$ & 16, 10 & 18,  6 & 0.270 & 0.257 & 22.95 & 15.08 \\
\isotope[82][42]{Mo}$_{40}$ & 18,  6 & 20,  2 & 0.255 & 0.248 & 15.08 &  6.59 \\
\isotope[84][42]{Mo}$_{42}$ & 12, 12 & 14,  8 & 0.245 & 0.228 & 30.00 & 21.79 \\
\hline
\end{tabular}

\end{table}

\begin{table}

\begin{tabular}{ r   r  r r c c r  }
\hline
nucleus  & hw & nhw & $\beta_{hw}$ & $\beta_{nhw}$ & $\gamma_{hw}$ & $\gamma_{nhw}$  \\ 
        
\hline

\isotope[74][44]{Ru}$_{30}$ &  8,  8 & 10,  4 & 0.177 & 0.161 & 30.00 & 17.78 \\
\isotope[76][44]{Ru}$_{32}$ & 10, 10 & 12,  6 & 0.214 & 0.198 & 30.00 & 20.17 \\
\isotope[78][44]{Ru}$_{34}$ & 14,  8 & 16,  4 & 0.234 & 0.223 & 21.79 & 12.52 \\
\isotope[80][44]{Ru}$_{36}$ & 12, 12 & 14,  8 & 0.249 & 0.232 & 30.00 & 21.79 \\
\isotope[82][44]{Ru}$_{38}$ & 12, 12 & 14,  8 & 0.247 & 0.230 & 30.00 & 21.79 \\
\isotope[84][44]{Ru}$_{40}$ & 14,  8 & 16,  4 & 0.228 & 0.217 & 21.79 & 12.52 \\
\isotope[86][44]{Ru}$_{42}$ &  8, 14 & 10, 10 & 0.227 & 0.206 & 38.21 & 30.00 \\
\hline
\end{tabular}

\end{table}

\begin{table}

\begin{tabular}{ r   r  r r c c r  }
\hline
nucleus  & hw & nhw & $\beta_{hw}$ & $\beta_{nhw}$ & $\gamma_{hw}$ & $\gamma_{nhw}$  \\ 
        
\hline

\isotope[76][46]{Pd}$_{30}$ &  6,  6 &  8,  2 & 0.136 & 0.122 & 30.00 & 13.90 \\
\isotope[78][46]{Pd}$_{32}$ &  8,  8 & 10,  4 & 0.174 & 0.158 & 30.00 & 17.78 \\
\isotope[80][46]{Pd}$_{34}$ & 12,  6 & 14,  2 & 0.194 & 0.185 & 20.17 &  8.95 \\
\isotope[82][46]{Pd}$_{36}$ & 10, 10 & 12,  6 & 0.209 & 0.193 & 30.00 & 20.17 \\
\isotope[84][46]{Pd}$_{38}$ & 10, 10 & 12,  6 & 0.207 & 0.191 & 30.00 & 20.17 \\
\isotope[86][46]{Pd}$_{40}$ & 12,  6 & 14,  2 & 0.190 & 0.180 & 20.17 &  8.95 \\
\isotope[88][46]{Pd}$_{42}$ &  6, 12 &  8,  8 & 0.188 & 0.167 & 39.83 & 30.00 \\
\hline
\end{tabular}

\end{table}


\begin{table}

\begin{tabular}{ r   r  r r c c r  }
\hline
nucleus  & hw & nhw & $\beta_{hw}$ & $\beta_{nhw}$ & $\gamma_{hw}$ & $\gamma_{nhw}$  \\ 
        
\hline

\isotope[78][48]{Cd}$_{30}$ &  6,  0 &  2,  2 & 0.084 & 0.058 &  6.59 & 30.00 \\
\isotope[80][48]{Cd}$_{32}$ &  8,  2 &  4,  4 & 0.120 & 0.096 & 13.90 & 30.00 \\
\isotope[82][48]{Cd}$_{34}$ & 12,  0 &  6,  6 & 0.148 & 0.133 &  3.67 & 30.00 \\
\isotope[84][48]{Cd}$_{36}$ & 10,  4 & 12,  0 & 0.154 & 0.147 & 17.78 &  3.67 \\
\isotope[86][48]{Cd}$_{38}$ & 10,  4 & 12,  0 & 0.153 & 0.146 & 17.78 &  3.67 \\
\isotope[88][48]{Cd}$_{40}$ & 12,  0 &  4, 10 & 0.145 & 0.152 &  3.67 & 42.22 \\
\isotope[90][48]{Cd}$_{42}$ &  6,  6 &  8,  2 & 0.129 & 0.115 & 30.00 & 13.90 \\
\hline
\end{tabular}

\end{table}

\section*{Appendix B}

Same as Appendix A, but for nuclei in the $Z=50$-82, $N=50$-184 region.

\begin{table}

\begin{tabular}{ r   r  r r c c r  }
\hline
nucleus  & hw & nhw & $\beta_{hw}$ & $\beta_{nhw}$ & $\gamma_{hw}$ & $\gamma_{nhw}$  \\ 
        
\hline

\isotope[104][52]{Te}$_{52}$ & 16,  0 & 12,  2 & 0.122 & 0.103 &  2.83 & 10.16 \\
\isotope[106][52]{Te}$_{54}$ & 20,  2 & 16,  4 & 0.157 & 0.138 &  6.59 & 12.52 \\
\isotope[108][52]{Te}$_{56}$ & 26,  0 & 20,  6 & 0.189 & 0.173 &  1.80 & 13.90 \\
\isotope[110][52]{Te}$_{58}$ & 26,  4 & 28,  0 & 0.204 & 0.202 &  8.35 &  1.68 \\
\isotope[112][52]{Te}$_{60}$ & 28,  4 & 30,  0 & 0.216 & 0.214 &  7.83 &  1.58 \\
\isotope[114][52]{Te}$_{62}$ & 32,  0 & 24, 10 & 0.226 & 0.216 &  1.48 & 17.35 \\
\isotope[116][52]{Te}$_{64}$ & 28,  6 & 30,  2 & 0.222 & 0.219 & 10.57 &  4.57 \\
\hline
\end{tabular}

\end{table}

\begin{table}

\begin{tabular}{ r   r  r r c c r  }
\hline
nucleus  & hw & nhw & $\beta_{hw}$ & $\beta_{nhw}$ & $\gamma_{hw}$ & $\gamma_{nhw}$  \\ 
        
\hline

\isotope[106][54]{Xe}$_{52}$ & 20,  2 & 16,  4 & 0.157 & 0.138 &  6.59 & 12.52 \\
\isotope[108][54]{Xe}$_{54}$ & 24,  4 & 20,  6 & 0.191 & 0.173 &  8.95 & 13.90 \\
\isotope[110][54]{Xe}$_{56}$ & 30,  2 & 24,  8 & 0.223 & 0.208 &  4.57 & 14.80 \\
\isotope[112][54]{Xe}$_{58}$ & 30,  6 & 32,  2 & 0.238 & 0.235 &  9.97 &  4.31 \\
\isotope[114][54]{Xe}$_{60}$ & 32,  6 & 34,  2 & 0.250 & 0.247 &  9.43 &  4.07 \\
\isotope[116][54]{Xe}$_{62}$ & 36,  2 & 28, 12 & 0.259 & 0.250 &  3.86 & 17.60 \\
\isotope[118][54]{Xe}$_{64}$ & 32,  8 & 34,  4 & 0.256 & 0.252 & 11.74 &  6.59 \\
\hline
\end{tabular}

\end{table}   

\begin{table}

\begin{tabular}{ r   r  r r c c r  }
\hline
nucleus  & hw & nhw & $\beta_{hw}$ & $\beta_{nhw}$ & $\gamma_{hw}$ & $\gamma_{nhw}$  \\ 
        
\hline   

\isotope[108][56]{Ba}$_{52}$ & 26,  0 & 20,  6 & 0.189 & 0.173 &  1.80 & 13.90 \\
\isotope[110][56]{Ba}$_{54}$ & 30,  2 & 24,  8 & 0.223 & 0.208 &  4.57 & 14.80 \\
\isotope[112][56]{Ba}$_{56}$ & 36,  0 & 30,  6 & 0.255 & 0.238 &  1.32 &  9.97 \\
\isotope[114][56]{Ba}$_{58}$ & 36,  4 & 38,  0 & 0.268 & 0.267 &  6.26 &  1.26 \\
\isotope[116][56]{Ba}$_{60}$ & 38,  4 & 40,  0 & 0.280 & 0.279 &  5.96 &  1.20 \\
\isotope[118][56]{Ba}$_{62}$ & 42,  0 & 34, 10 & 0.290 & 0.278 &  1.14 & 13.24 \\
\isotope[120][56]{Ba}$_{64}$ & 38,  6 & 40,  2 & 0.285 & 0.283 &  8.12 &  3.50 \\
\hline
\end{tabular}

\end{table}   

\begin{table} 

\begin{tabular}{ r   r  r r c c r  }
\hline
nucleus  & hw & nhw & $\beta_{hw}$ & $\beta_{nhw}$ & $\gamma_{hw}$ & $\gamma_{nhw}$  \\ 
        
\hline

\isotope[110][58]{Ce}$_{52}$ & 26,  4 & 28,  0 & 0.204 & 0.202 &  8.35 &  1.68 \\
\isotope[112][58]{Ce}$_{54}$ & 30,  6 & 32,  2 & 0.238 & 0.235 &  9.97 &  4.31 \\
\isotope[114][58]{Ce}$_{56}$ & 36,  4 & 38,  0 & 0.268 & 0.267 &  6.26 &  1.26 \\
\isotope[116][58]{Ce}$_{58}$ & 36,  8 & 38,  4 & 0.283 & 0.280 & 10.64 &  5.96 \\
\isotope[118][58]{Ce}$_{60}$ & 38,  8 & 40,  4 & 0.295 & 0.292 & 10.16 &  5.69 \\
\isotope[120][58]{Ce}$_{62}$ & 42,  4 & 44,  0 & 0.303 & 0.302 &  5.44 &  1.09 \\
\isotope[122][58]{Ce}$_{64}$ & 38, 10 & 40,  6 & 0.300 & 0.296 & 12.08 &  7.76 \\
\hline
\end{tabular}

\end{table}   
   
\begin{table}

\begin{tabular}{ r   r  r r c c r  }
\hline
nucleus  & hw & nhw & $\beta_{hw}$ & $\beta_{nhw}$ & $\gamma_{hw}$ & $\gamma_{nhw}$  \\ 
        
\hline    	

\isotope[112][60]{Nd}$_{52}$ & 28,  4 & 30,  0 & 0.216 & 0.214 &  7.83 &  1.58 \\
\isotope[114][60]{Nd}$_{54}$ & 32,  6 & 34,  2 & 0.250 & 0.247 &  9.43 &  4.07 \\
\isotope[116][60]{Nd}$_{56}$ & 38,  4 & 40,  0 & 0.280 & 0.279 &  5.96 &  1.20 \\
\isotope[118][60]{Nd}$_{58}$ & 38,  8 & 40,  4 & 0.295 & 0.292 & 10.16 &  5.69 \\
\isotope[120][60]{Nd}$_{60}$ & 40,  8 & 42,  4 & 0.306 & 0.303 &  9.72 &  5.44 \\
\isotope[122][60]{Nd}$_{62}$ & 44,  4 & 46,  0 & 0.315 & 0.314 &  5.21 &  1.05 \\
\isotope[124][60]{Nd}$_{64}$ & 40, 10 & 42,  6 & 0.312 & 0.308 & 11.58 &  7.43 \\
\hline
\end{tabular}

\end{table}   

\begin{table}

\begin{tabular}{ r   r  r r c c r  }
\hline
nucleus  & hw & nhw & $\beta_{hw}$ & $\beta_{nhw}$ & $\gamma_{hw}$ & $\gamma_{nhw}$  \\ 
        
\hline

\isotope[114][62]{Sm}$_{52}$ & 32,  0 & 24, 10 & 0.226 & 0.216 &  1.48 & 17.35 \\
\isotope[116][62]{Sm}$_{54}$ & 36,  2 & 28, 12 & 0.259 & 0.250 &  3.86 & 17.60 \\
\isotope[118][62]{Sm}$_{56}$ & 42,  0 & 34, 10 & 0.290 & 0.278 &  1.14 & 13.24 \\
\isotope[120][62]{Sm}$_{58}$ & 42,  4 & 44,  0 & 0.303 & 0.302 &  5.44 &  1.09 \\
\isotope[122][62]{Sm}$_{60}$ & 44,  4 & 46,  0 & 0.315 & 0.314 &  5.21 &  1.05 \\
\isotope[124][62]{Sm}$_{62}$ & 48,  0 & 40, 10 & 0.325 & 0.312 &  1.00 & 11.58 \\
\isotope[126][62]{Sm}$_{64}$ & 44,  6 & 46,  2 & 0.319 & 0.317 &  7.13 &  3.07 \\
\hline
\end{tabular}

\end{table}

\begin{table}

\begin{tabular}{ r   r  r r c c r  }
\hline
nucleus  & hw & nhw & $\beta_{hw}$ & $\beta_{nhw}$ & $\gamma_{hw}$ & $\gamma_{nhw}$  \\ 
        
\hline

\isotope[116][64]{Gd}$_{52}$ & 28,  6 & 30,  2 & 0.222 & 0.219 & 10.57 &  4.57 \\
\isotope[118][64]{Gd}$_{54}$ & 32,  8 & 34,  4 & 0.256 & 0.252 & 11.74 &  6.59 \\
\isotope[120][64]{Gd}$_{56}$ & 38,  6 & 40,  2 & 0.285 & 0.283 &  8.12 &  3.50 \\
\isotope[122][64]{Gd}$_{58}$ & 38, 10 & 40,  6 & 0.300 & 0.296 & 12.08 &  7.76 \\
\isotope[124][64]{Gd}$_{60}$ & 40, 10 & 42,  6 & 0.312 & 0.308 & 11.58 &  7.43 \\
\isotope[126][64]{Gd}$_{62}$ & 44,  6 & 46,  2 & 0.319 & 0.317 &  7.13 &  3.07 \\
\isotope[128][64]{Gd}$_{64}$ & 40, 12 & 42,  8 & 0.317 & 0.313 & 13.33 &  9.32 \\
\hline
\end{tabular}

\end{table}

\begin{table}

\begin{tabular}{ r   r  r r c c r  }
\hline
nucleus  & hw & nhw & $\beta_{hw}$ & $\beta_{nhw}$ & $\gamma_{hw}$ & $\gamma_{nhw}$  \\ 
        
\hline

\isotope[118][66]{Dy}$_{52}$ & 26,  8 & 28,  4 & 0.217 & 0.212 & 13.90 &  7.83 \\
\isotope[120][66]{Dy}$_{54}$ & 30, 10 & 32,  6 & 0.250 & 0.246 & 14.63 &  9.43 \\
\isotope[122][66]{Dy}$_{56}$ & 36,  8 & 38,  4 & 0.279 & 0.275 & 10.64 &  5.96 \\
\isotope[124][66]{Dy}$_{58}$ & 36, 12 & 38,  8 & 0.295 & 0.290 & 14.51 & 10.16 \\
\isotope[126][66]{Dy}$_{60}$ & 38, 12 & 40,  8 & 0.306 & 0.301 & 13.90 &  9.72 \\
\isotope[128][66]{Dy}$_{62}$ & 42,  8 & 44,  4 & 0.313 & 0.310 &  9.32 &  5.21 \\
\isotope[130][66]{Dy}$_{64}$ & 38, 14 & 40, 10 & 0.312 & 0.307 & 15.61 & 11.58 \\
\hline
\end{tabular}

\end{table}   
   
\begin{table}

\begin{tabular}{ r   r  r r c c r  }
\hline
nucleus  & hw & nhw & $\beta_{hw}$ & $\beta_{nhw}$ & $\gamma_{hw}$ & $\gamma_{nhw}$  \\ 
        
\hline                             

\isotope[120][68]{Er}$_{52}$ & 26,  6 & 28,  2 & 0.206 & 0.203 & 11.24 &  4.87 \\
\isotope[122][68]{Er}$_{54}$ & 30,  8 & 32,  4 & 0.240 & 0.236 & 12.38 &  6.95 \\
\isotope[124][68]{Er}$_{56}$ & 36,  6 & 38,  2 & 0.269 & 0.266 &  8.51 &  3.67 \\
\isotope[126][68]{Er}$_{58}$ & 36, 10 & 38,  6 & 0.284 & 0.280 & 12.63 &  8.12 \\
\isotope[128][68]{Er}$_{60}$ & 38, 10 & 40,  6 & 0.296 & 0.292 & 12.08 &  7.76 \\
\isotope[130][68]{Er}$_{62}$ & 42,  6 & 44,  2 & 0.303 & 0.301 &  7.43 &  3.20 \\
\isotope[132][68]{Er}$_{64}$ & 38, 12 & 40,  8 & 0.301 & 0.297 & 13.90 &  9.72 \\
\hline
\end{tabular}

\end{table}

\begin{table}

\begin{tabular}{ r   r  r r c c r  }
\hline
nucleus  & hw & nhw & $\beta_{hw}$ & $\beta_{nhw}$ & $\gamma_{hw}$ & $\gamma_{nhw}$  \\ 
        
\hline

\isotope[122][70]{Yb}$_{52}$ & 28,  0 & 18, 14 & 0.195 & 0.195 &  1.68 & 26.11 \\
\isotope[124][70]{Yb}$_{54}$ & 32,  2 & 22, 16 & 0.227 & 0.228 &  4.31 & 25.05 \\
\isotope[126][70]{Yb}$_{56}$ & 38,  0 & 28, 14 & 0.258 & 0.253 &  1.26 & 19.59 \\
\isotope[128][70]{Yb}$_{58}$ & 38,  4 & 40,  0 & 0.271 & 0.270 &  5.96 &  1.20 \\
\isotope[130][70]{Yb}$_{60}$ & 40,  4 & 42,  0 & 0.282 & 0.281 &  5.69 &  1.14 \\
\isotope[132][70]{Yb}$_{62}$ & 44,  0 & 34, 14 & 0.293 & 0.286 &  1.09 & 17.00 \\
\isotope[134][70]{Yb}$_{64}$ & 40,  6 & 42,  2 & 0.287 & 0.285 &  7.76 &  3.34 \\
\hline
\end{tabular}

\end{table}      
   
\begin{table}

\begin{tabular}{ r   r  r r c c r  }
\hline
nucleus  & hw & nhw & $\beta_{hw}$ & $\beta_{nhw}$ & $\gamma_{hw}$ & $\gamma_{nhw}$  \\ 
        
\hline   

\isotope[124][72]{Hf}$_{52}$ & 20,  8 & 22,  4 & 0.175 & 0.170 & 17.00 &  9.64 \\
\isotope[126][72]{Hf}$_{54}$ & 24, 10 & 26,  6 & 0.209 & 0.203 & 17.35 & 11.24 \\
\isotope[128][72]{Hf}$_{56}$ & 30,  8 & 32,  4 & 0.236 & 0.232 & 12.38 &  6.95 \\
\isotope[130][72]{Hf}$_{58}$ & 30, 12 & 32,  8 & 0.253 & 0.247 & 16.71 & 11.74 \\
\isotope[132][72]{Hf}$_{60}$ & 32, 12 & 34,  8 & 0.264 & 0.259 & 15.91 & 11.16 \\
\isotope[134][72]{Hf}$_{62}$ & 36,  8 & 38,  4 & 0.270 & 0.267 & 10.64 &  5.96 \\
\isotope[136][72]{Hf}$_{64}$ & 32, 14 & 34, 10 & 0.271 & 0.265 & 17.78 & 13.24 \\
\hline
\end{tabular}

\end{table}      

\begin{table}

\begin{tabular}{ r   r  r r c c r  }
\hline
nucleus  & hw & nhw & $\beta_{hw}$ & $\beta_{nhw}$ & $\gamma_{hw}$ & $\gamma_{nhw}$  \\ 
        
\hline

\isotope[126][74]{W}$_{52}$ & 14, 12 & 16,  8 & 0.158 & 0.149 & 27.64 & 19.93 \\
\isotope[128][74]{W}$_{54}$ & 18, 14 & 20, 10 & 0.192 & 0.183 & 26.11 & 19.77 \\
\isotope[130][74]{W}$_{56}$ & 24, 12 & 26,  8 & 0.216 & 0.210 & 19.67 & 13.90 \\
\isotope[132][74]{W}$_{58}$ & 24, 16 & 26, 12 & 0.235 & 0.227 & 23.72 & 18.58 \\
\isotope[134][74]{W}$_{60}$ & 26, 16 & 28, 12 & 0.246 & 0.238 & 22.52 & 17.60 \\
\isotope[136][74]{W}$_{62}$ & 30, 12 & 32,  8 & 0.249 & 0.244 & 16.71 & 11.74 \\
\isotope[138][74]{W}$_{64}$ & 26, 18 & 28, 14 & 0.254 & 0.245 & 24.27 & 19.59 \\
\hline
\end{tabular}

\end{table}         
   
\begin{table}

\begin{tabular}{ r   r  r r c c r  }
\hline
nucleus  & hw & nhw & $\beta_{hw}$ & $\beta_{nhw}$ & $\gamma_{hw}$ & $\gamma_{nhw}$  \\ 
        
\hline  

\isotope[128][76]{Os}$_{52}$ & 10, 12 & 12,  8 & 0.135 & 0.124 & 32.75 & 24.01 \\
\isotope[130][76]{Os}$_{54}$ & 14, 14 & 16, 10 & 0.168 & 0.158 & 30.00 & 22.95 \\
\isotope[132][76]{Os}$_{56}$ & 20, 12 & 22,  8 & 0.191 & 0.184 & 22.26 & 15.82 \\
\isotope[134][76]{Os}$_{58}$ & 20, 16 & 22, 12 & 0.211 & 0.202 & 26.52 & 20.89 \\
\isotope[136][76]{Os}$_{60}$ & 22, 16 & 24, 12 & 0.221 & 0.213 & 25.05 & 19.67 \\
\isotope[138][76]{Os}$_{62}$ & 26, 12 & 28,  8 & 0.224 & 0.218 & 18.58 & 13.10 \\
\isotope[140][76]{Os}$_{64}$ & 22, 18 & 24, 14 & 0.230 & 0.221 & 26.85 & 21.79 \\
\hline
\end{tabular}

\end{table}         

\begin{table}

\begin{tabular}{ r   r  r r c c r  }
\hline
nucleus  & hw & nhw & $\beta_{hw}$ & $\beta_{nhw}$ & $\gamma_{hw}$ & $\gamma_{nhw}$  \\ 
        
\hline

\isotope[130][78]{Pt}$_{52}$ &  8,  8 & 10,  4 & 0.101 & 0.092 & 30.00 & 17.78 \\
\isotope[132][78]{Pt}$_{54}$ & 12, 10 & 14,  6 & 0.134 & 0.125 & 27.25 & 18.14 \\
\isotope[134][78]{Pt}$_{56}$ & 18,  8 & 20,  4 & 0.158 & 0.153 & 18.35 & 10.44 \\
\isotope[136][78]{Pt}$_{58}$ & 18, 12 & 20,  8 & 0.177 & 0.170 & 23.82 & 17.00 \\
\isotope[138][78]{Pt}$_{60}$ & 20, 12 & 22,  8 & 0.188 & 0.181 & 22.26 & 15.82 \\
\isotope[140][78]{Pt}$_{62}$ & 24,  8 & 26,  4 & 0.192 & 0.188 & 14.80 &  8.35 \\
\isotope[142][78]{Pt}$_{64}$ & 20, 14 & 22, 10 & 0.197 & 0.189 & 24.50 & 18.48 \\
\hline
\end{tabular}

\end{table}            
   
\begin{table} 

\begin{tabular}{ r   r  r r c c r  }
\hline
nucleus  & hw & nhw & $\beta_{hw}$ & $\beta_{nhw}$ & $\gamma_{hw}$ & $\gamma_{nhw}$  \\ 
        
\hline   

\isotope[132][80]{Hg}$_{52}$ &  8,  0 &  4,  2 & 0.061 & 0.045 &  5.21 & 21.79 \\
\isotope[134][80]{Hg}$_{54}$ & 12,  2 &  8,  4 & 0.094 & 0.079 & 10.16 & 20.63 \\
\isotope[136][80]{Hg}$_{56}$ & 18,  0 & 12,  6 & 0.124 & 0.112 &  2.54 & 20.17 \\
\isotope[138][80]{Hg}$_{58}$ & 18,  4 & 20,  0 & 0.139 & 0.136 & 11.39 &  2.31 \\
\isotope[140][80]{Hg}$_{60}$ & 20,  4 & 22,  0 & 0.151 & 0.148 & 10.44 &  2.11\\
\isotope[142][80]{Hg}$_{62}$ & 24,  0 & 16, 10 & 0.160 & 0.153 &  1.95 & 22.95 \\
\isotope[144][80]{Hg}$_{64}$ & 20,  6 & 22,  2 & 0.158 & 0.154 & 13.90 &  6.05 \\
\hline
\end{tabular}

\end{table}

\section*{Appendix C}

In this Appendix a proof is provided for the occurrence of $\mu=0$ in the hw irreps of any U(N) for particle numbers $M=2$, 6, 12, 20, 30, 42, 56. 

The unitary algebras U(($\eta+1)(\eta+2)/2$), $\eta=1$, 2, 3, \dots are known to possess SU(3) subalgebras \cite{Bonatsos1986}, the single particle states of which can be described in terms of the numbers $n_z$, $n_x$, $n_y$ of oscillator quanta along the $z$, $x$, and $y$  axes as $(n_z, n_x, n_y)$, with 
$n_z+n_x+n_y=\eta$ \cite{Kota2020}.  
The single-particle states can be seen in Table IX. Each line of the table contains all possible single-particle states with $n_z=\eta-r$, with $r=0$, 1, 2, \dots, where $r$ is the number of quanta removed from the $z$-axis. We see that each line contains $r+1$ states, i.e., it can accommodate $2(r+1)$ particles. The number of particles which can be accommodated in each line is shown in the column $n_r$, while the total number of particles  up to a given line is shown in the column $N_r$. 

For the hw irrep $(\lambda_H, \mu_H)$ of SU(3) it is known that the Elliott quantum numbers \cite{Elliott1958a,Harvey1968} are given by 
\cite{Harvey1968,Martinou2021b}
\begin{equation}
\lambda_H = \sum_i n_{z,i} - \sum_i n_{x,i}, \qquad \mu_H = \sum_i n_{x,i} - \sum_i n_{y,i},
\end{equation}
where $i$ regards the participating single-particle orbitals, each occupied by two particles. 
Using these equations one can easily determine $\lambda_H$ and $\mu_H$ from the entries of Table IX.

For 2 particles, from the first line of Table IX one obtains for the hw irrep $(2\eta,0)$. 

For 6 particles, from the first two lines of Table IX one obtains $(6(\eta-1),0)$. 

For 12 particles. the first three lines of Table IX provide $(12(\eta-2),0)$.

For 20 particles, the first four lines of Table IX provide $(20(\eta-3),0)$.

For 30 particles, the first five lines of Table IX provide $(30(\eta-4),0)$.

For 42 particles, the first six lines of Table IX provide $(42(\eta-5),0)$.

For 56 particles, the first seven lines of Table IX provide $(56(\eta-6),0)$.

These results provide all hw irreps with $\mu=0$ appearing in Table VI. 

For example, for U(21) ($\eta=5$), the hw irreps for 2, 6, 12, 20, 30, 42 particles are (10,0), (24,0), (36,0), (40,0), (30,0), (0,0) respectively, 
in agreement with Table VI.  


\newpage
\setcounter{table}{0}



\begin{table}

\caption{Highest weight (hw) irreducible representations of SU(3) for $M$ nucleons occurring in the reduction U(6)$\supset$SU(3), calculated using the code of Ref. \cite{Draayer1989}. See Sec. \ref{hw} for further discussion.
}
\begin{tabular}{ r r r r r   }
\hline
$M$&     &     &         &     \\
      \hline
      
 2 & 4,0 & 0,2 &         &     \\
 4 & 4,2 & 0,4 & 2,0     &     \\ 
 6 & 6,0 & 0,6 & 2,2$^2$ & 0,0 \\
 8 & 2,4 & 4,0 & 0,2     &     \\   
10 & 0,4 & 2,0 &         &     \\
12 & 0,0 &     &         &     \\

\hline
\end{tabular}

\end{table}


\begin{table}

\caption{Highest weight (hw) irreducible representations of SU(3) for $M$ nucleons occurring in the reduction U(10)$\supset$SU(3), calculated using the code of Ref. \cite{Draayer1989}. See Sec. \ref{hw} for further discussion.
}
\begin{tabular}{ r r r r r r r r r r r r r   }
\hline
$M$ &     &         &         &         &         &         &            &            &         &         &            &     \\
      \hline
      
 2 & 6,0  & 2,2     &         &         &         &         &            &            &         &         &            &     \\  
 4 & 8,2  & 4,4$^2$ & 6,0     & 0,6     & 2,2$^2$ &         &            &            &         &         &            &     \\
 6 & 12,0 & 6,6     & 8,2$^3$ & 2,8     & 4,4$^5$ & 6,0$^4$ & 0,6$^3$    & 2,2$^5$    & 0,0$^2$ &         &            &     \\ 
 8 & 10,4 & 12,0    & 6,6$^3$ & 8,2$^5$ & 0,12    & 2,8$^4$ & 4,4$^{10}$ & 6,0$^6$    & 0,6$^5$ & 2,2$^8$ & 0,0$^2$    &     \\
10 & 10,4 & 12,0    & 4,10    & 6,6$^4$ & 8,2$^6$ & 0,12    & 2,8$^6$    & 4,4$^{14}$ & 6,0$^6$ & 0,6$^6$ & 2,2$^{11}$ & 0,0 \\
12 & 12,0 & 4,10    & 6,6$^3$ & 8,2$^4$ & 0,12    & 2,8$^5$ & 4,4$^{10}$ & 6,0$^5$    & 0,6$^6$ & 2,2$^8$ & 0,0$^2$    &     \\ 
14 & 6,6  & 8,2     & 0,12    & 2,8$^3$ & 4,4$^5$ & 6,0$^3$ & 0,6$^4$    & 2,2$^5$    & 0,0$^2$ &         &            &     \\
16 & 2,8  & 4,4$^2$ & 6,0     & 0,6     & 2,2$^2$ &         &            &            &         &         &            &     \\
18 & 0,6  & 2,2     &         &         &         &         &            &            &         &         &            &     \\
20 & 0,0  &         &         &         &         &         &            &            &         &         &            &     \\ 
 
\hline
\end{tabular}

\end{table}


\begin{table}

\caption{Highest weight (hw) irreducible representations of SU(3) for $M$ nucleons occurring in the reduction U(15)$\supset$SU(3), calculated using the code of Ref. \cite{Draayer1989}. See Sec. \ref{hw} for further discussion. 
}
\begin{tabular}{ r r r r r r r r r   }
\hline
$M$ &     &          &           &           &             &             &             &             \\    
      \hline
      
 2 & 8,0  & 4,2      & 0,4       &           &             &             &             &             \\ 
 4 & 12,2 & 8,4$^2$  & 10,0      & 4,6$^2$   & 6,2$^3$     & 0,8$^2$     & 2,4$^3$     & 4,0$^2$     \\ 
 6 & 18,0 & 12,6     & 14,2$^3$  & 8,8$^2$   & 10,4$^8$     & 12,0$^5$    & 4,10$^2$    & 6,6$^{14}$  \\ 
 8 & 18,4 & 20,0     & 14,6$^4$  & 16,2$^6$  & 8,12        & 10,8$^{11}$ & 12,4$^{24}$ & 14,0$^{11}$ \\ 
10 & 20,4 & 22,0     & 14,10     & 16,6$^6$  & 18,2$^8$    & 10,12$^4$   & 12,8$^{23}$ & 14,4$^{43}$ \\ 
12 & 24,0 & 16,10    & 18,6$^4$  & 20,2$^5$  & 12,12$^5$   & 14,8$^{23}$ & 16,4$^{37}$ & 18,0$^{15}$ \\ 
14 & 20,6 & 22,2     & 14,12$^2$ & 16,8$^9$  & 18,4$^{14}$ & 20,0$^6$    & 8,18        & 10,14$^9$   \\ 
16 & 18,8 & 20,4$^2$ &      22,0 & 12,14$^2$ & 14,10$^9$   & 16,6$^{20}$ & 18,2$^{18}$ & 6,20        \\ 
18 & 18,6 & 20,2     & 10,16     & 12,12$^5$ & 14,8$^{13}$ & 16,4$^{17}$ & 18,0$^7$    & 6,18$^4$       \\ 
20 & 20,0 & 10,14    & 12,10$^4$ & 14,6$^7$  & 16,2$^6$    & 4,20        & 6,16$^6$    & 8,12$^{23}$ \\ 
22 & 12,8 & 14,4     & 16,0      & 4,18      & 6,14$^4$    & 8,10$^{11}$ & 10,6$^{16}$ & 12,2$^{12}$ \\ 
24 & 6,12 & 8,8$^2$  & 10,4$^2$   & 12,0$^2$  & 0,18        & 2,14$^3$    & 4,10$^8$    & 6,6$^{14}$  \\ 
26 & 2,12 & 4,8$^2$  & 6,4$^2$   & 8,0$^2$   & 0,10        & 2,6$^3$     & 4,2$^3$     & 0,4$^2$     \\ 
28 & 0,8  & 2,4      & 4,0       &           &             &             &             &             \\ 
30 & 0,0  &          &           &           &             &             &             &             \\ 

\hline
\end{tabular}

\end{table}


\begin{table}

\caption{Highest weight (hw) irreducible representations of SU(3) for $M$ nucleons occurring in the reduction U(21)$\supset$SU(3), calculated using the code of Ref. \cite{Draayer1989}. See Sec. \ref{hw} for further discussion.
}
\begin{tabular}{ r r r r r r r r r   }
\hline
$M$ &       &           &           &             &              &              &              &             \\ 
      \hline
      
  2 & 10,0  & 6,2       & 2,4       &             &              &              &              &              \\
  4 & 16,2  & 12,4$^2$  & 14,0      & 8,6$^3$     & 10,2$^3$     & 4,8$^3$      & 6,4$^5$      & 8,0$^2$      \\
  6 & 24,0  & 18,6      & 20,2$^3$  & 14,8$^2$    & 16,4$^9$     & 18,0$^5$     & 10,10$^3$    & 12,6$^{20}$  \\ 
  8 & 26,4  & 28,0      & 22,6$^4$  & 24,2$^6$    & 16,12        & 18,8$^{14}$  & 20,4$^{28}$  & 22,0$^{12}$  \\
 10 & 30,4  & 32,0      & 24,10     & 26,6$^6$    & 28,2$^8$     & 20,12$^5$    & 22,8$^{29}$  & 24,4$^{52}$  \\
 12 & 36,0  & 28,10     & 30,6$^4$  & 32,2$^5$    & 24,12$^6$    & 26,8$^{28}$  & 28,4$^{44}$  & 30,0$^{17}$  \\
 14 & 34,6  & 36,2      & 28,12$^2$ & 30,8$^{10}$ & 32,4$^{15}$  & 34,0$^6$     & 22,18        & 24,14$^{14}$ \\
 16 & 34,8  & 36,4$^2$  & 38,0      & 28,14$^2$   & 30,10$^{11}$ & 32,6$^{24}$  & 34,2$^{21}$  & 22,20        \\
 18 & 36,6  & 38,2      & 28,16     & 30,12$^6$   & 32,8$^{16}$  & 34,4$^{21}$  & 36,0$^8$     & 24,18$^{10}$ \\
 20 & 40,0  & 30,14     & 32,10$^4$ & 34,6$^8$    & 36,2$^7$     & 24,20$^2$    & 26,16$^{16}$ & 28,12$^{60}$ \\
 22 & 34,8  & 36,4      & 38,0      & 26,18       & 28,14$^7$    & 30,10$^{21}$ & 32,6$^{31}$  & 34,2$^{22}$  \\
 24 & 30,12 & 32,8$^2$  & 34,4$^2$  & 36,0$^2$    & 24,18$^5$    & 26,14$^{20}$ & 28,10$^{46}$ & 30,6$^{63}$  \\
 26 & 28,12 & 30,8$^2$  & 32,4$^2$  & 34,0$^2$    & 20,22        & 22,18$^7$    & 24,14$^{25}$ & 26,10$^{52}$ \\
 28 & 28,8  & 30,4      & 32,0      & 18,22       & 20,18$^5$    & 22,14$^{15}$ & 24,10$^{31}$ & 26,6$^{39}$  \\
 30 & 30,0  & 18,18     & 20,14$^4$ & 22,10$^8$   & 24,6$^{11}$  & 26,2$^8$     & 10,28        & 12,24$^6$    \\
 32 & 20,10 & 22,6      & 24,2      & 10,24       & 12,20$^5$    & 14,16$^{14}$ & 16,12$^{28}$ & 18,8$^{38}$  \\
 34 & 12,16 & 14,12$^2$ & 16,8$^3$  & 18,4$^3$    & 20,0$^2$     & 4,26         & 6,22$^4$     & 8,18$^{14}$  \\
 36 & 6,18  & 8,14$^2$  & 10,10$^3$ & 12,6$^5$    & 14,2$^3$     & 0,24         & 2,20$^3$     & 4,16$^9$     \\
 38 & 2,16  & 4,12$^2$  & 6,8$^3$   & 8,4$^3$     & 10,0$^2$     & 0,14         & 2,10$^3$     & 4,6$^5$      \\
 40 & 0,10  & 2,6       & 4,2       &             &              &              &              &              \\
 42 & 0,0   &           &           &             &              &              &              &              \\

\hline
\end{tabular}

\end{table}


\begin{table}

\caption{Highest weight (hw) irreducible representations of SU(3) for $M$ nucleons occurring in the reduction U(28)$\supset$SU(3), calculated using the code of Ref. \cite{Draayer1989}. See Sec. \ref{hw} for further discussion.
}
\begin{tabular}{ r r r r r r r r r   }
\hline
$M$&       &           &           &             &              &              &              &             \\ 
      \hline
      
 2 & 12,0  & 8,2       & 4,4       & 0,6         &              &              &              &              \\
 4 & 20,2  & 16,4$^2$  & 18,0      & 12,6$^3$    & 14,2$^3$     & 8,8$^4$      & 10,4$^6$     & 12,0$^2$     \\ 
 6 & 30,0  & 24,6      & 26,2$^3$  & 20,8$^2$    & 22,4$^9$     & 24,0$^5$     & 16,10$^4$    & 18,6$^{23}$  \\
 8 & 34,4  & 36,0      & 30,6$^4$  & 32,2$^6$    & 24,12        & 26,8$^{15}$   & 28,4$^{29}$  & 30,0$^{12}$  \\
10 & 40,4  & 42,0      & 34,10     & 36,6$^6$    & 38,2$^8$     & 30,12$^5$    & 32,8$^{31}$  & 34,4$^{54}$  \\
12 & 48,0  & 40,10     & 42,6$^4$  & 44,2$^5$    & 36,12$^6$    & 38,8$^{29}$  & 40,4$^{45}$  & 42,0$^{17}$  \\
14 & 48,6  & 50,2      & 42,12$^2$ & 44,8$^{10}$ & 46,4$^{15}$  & 48,0$^6$     & 36,18        & 38,14$^{15}$ \\
16 & 50,8  & 52,4$^2$  & 54,0      & 44,14$^2$   & 46,10$^{11}$ & 48,6$^{24}$  & 50,2$^{21}$  & 38,20        \\
18 & 54,6  & 56,2      & 46,16     & 48,12$^6$   & 50,8$^{16}$  & 52,4$^{21}$  & 54,0$^8$     & 42,18$^{11}$ \\
20 & 60,0  & 50,14     & 52,10$^4$ & 54,6$^8$    & 56,2$^7$     & 44,20$^2$    & 46,16$^{17}$ & 48,12$^{66}$ \\
22 & 56,8  & 58,4      & 60,0      & 48,18       & 50,14$^7$    & 52,10$^{22}$ & 54,6$^{32}$  & 56,2$^{23}$  \\
24 & 54,12 & 56,8$^2$  & 58,4$^2$  & 60,0$^2$    & 48,18$^5$    & 50,14$^{22}$ & 52,10$^{51}$ & 54,6$^{69}$  \\
26 & 54,12 & 56,8$^2$  & 58,4$^2$  & 60,0$^2$    & 46,22        & 48,18$^8$    & 50,14$^{29}$ & 52,10$^{61}$ \\
28 & 56,8  & 58,4      & 60,0      & 46,22       & 48,18$^5$    & 50,14$^{17}$ & 52,10$^{36}$ & 54,6$^{45}$  \\
30 & 60,0  & 48,18     & 50,14$^4$ & 52,10$^9$   & 54,6$^{12}$  & 56,2$^9$     & 40,28        & 42,24$^{10}$ \\
32 & 52,10 & 54,6      & 56,2      & 42,24       & 44,20$^6$    & 46,16$^{20}$ & 48,12$^{43}$ & 50,8$^{60}$  \\
34 & 46,16 & 48,12$^2$ & 50,8$^3$  & 52,4$^3$    & 54,0$^2$     & 38,26$^2$    & 40,22$^{13}$ & 42,18$^{44}$ \\
36 & 42,18 & 44,14$^2$ & 46,10$^3$ & 48,6$^5$    & 50,2$^3$     & 34,28$^2$    & 36,24$^{13}$ & 38,20$^{46}$ \\
38 & 40,16 & 42,12$^2$ & 44,8$^3$  & 46,4$^3$    & 48,0$^2$     & 30,30        & 32,26$^6$    & 34,22$^{23}$ \\
40 & 40,10 & 42,6      & 44,2      & 28,28       & 30,24$^5$    & 32,20$^{16}$ & 34,16$^{37}$ & 36,12$^{62}$ \\
42 & 42,0  & 28,22     & 30,18$^4$ & 32,14$^9$   & 34,10$^{14}$ & 36,6$^{15}$  & 38,2$^{10}$  & 18,36        \\
44 & 30,12 & 32,8      & 34,4      & 36,0        & 18,30        & 20,26$^5$    & 22,22$^{15}$ & 24,18$^{35}$ \\
46 & 20,20 & 22,16$^2$ & 24,12$^3$ & 26,8$^4$    & 28,4$^4$     & 30,0$^2$     & 10,34        &  12,30$^5$   \\
48 & 12,24 & 14,20$^2$ & 16,16$^4$ & 18,12$^7$   & 20,8$^8$     & 22,4$^{6}$   & 24,0$^4$     &  4,34        \\
50 & 6,24  & 8,20$^2$  & 10,16$^4$ & 12,12$^7$   & 14,8$^8$     & 16,4$^6$     & 18,0$^4$     & 0,30         \\
52 & 2,20  & 4,16$^2$  & 6,12$^3$  & 8,8$^4$     & 10,4$^4$     & 12,0$^2$     & 0,18         & 2,14$^3$     \\
54 & 0,12  & 2,8       & 4,4       & 6,0         &              &              &              &              \\
56 & 0,0   &           &           &             &              &              &              &              \\

\hline
\end{tabular}

\end{table}


\begin{table*}

\caption{Highest weight (hw) irreducible representations of SU(3) and next highest weight (nhw) irreps of SU(3) for $M$ nucleons within the proxy-SU(3) scheme in the $sd$, $pf$, $sdg$, $pfh$, $sdgi$ shells having the overall symmetry  U(6), U(10), U(15), U(21), and U(28) respectively, calculated using the code UNTOU3 \cite{Draayer1989a}. The Elliott \cite{Elliott1958a} notation $(\lambda,\mu)$ is used for the SU(3) irreps.  
The corresponding shells of the shell model, within the proxy-SU(3) scheme \cite{Martinou2020} are also shown. Adapted from Ref. \cite{Bonatsos2024b}.  
 See Section \ref{hw} for further discussion. 
}
\begin{tabular}{ r r r r r r r r r r r  }
\hline
$M$  & U(6) & U(6) & U(10) & U(10) & U(15) & U(15) & U(21) & U(21) & U(28) & U(28) \\
     & $sd$ & $sd$ & $pf$  & $pf$  & $sdg$ & $sdg$ & $pfh$ & $pfh$ & $sdgi$ & $sdgi$ \\
     & 8-20 & 8-20 & 28-50 & 28-50 & 50-82 & 50-82 & 82-126 & 82-126 & 126-128 & 126-184 \\
     & hw   & nhw  & hw    & nhw   & hw    & nhw   & hw    & nhw   & hw    & nhw   \\

      \hline
 2 & 4,0 & 0,2 &  6,0 &  2,2 &  8,0 &   4,2 &  10,0 &   6,2 &  12,0 &   8,2 \\ 
 4 & 4,2 & 0,4 &  8,2 &  4,4 & 12,2 &   8,4 &  16,2 &  12,4 &  20,2 &  16,4 \\
 6 & 6,0 & 0,6 & 12.0 &  6,6 & 18,0 &  12,6 &  24,0 &  18,6 &  30,0 &  24,6 \\
 8 & 2,4 & 4,0 & 10,4 & 12,0 & 18,4 &  20,0 &  26,4 &  28,0 &  34,4 &  36,0 \\
10 & 0,4 & 2,0 & 10,4 & 12,0 & 20,4 &  22,0 &  30,4 &  32,0 &  40,4 &  42,0 \\
12 & 0,0 &     & 12,0 & 4,10 & 24,0 & 16,10 &  36,0 & 28,10 &  48,0 & 40,10 \\
14 &     &     &  6,6 &  8,2 & 20,6 &  22,2 &  34,6 &  36,2 &  48,6 &  50,2 \\
16 &     &     &  2,8 &  4,4 & 18,8 &  20,4 &  34,8 &  36,4 &  50,8 &  52,4 \\
18 &     &     &  0,6 &  2,2 & 18,6 &  20,2 &  36,6 &  38,2 &  54,6 &  56,2 \\
20 &     &     &  0,0 &      & 20,0 & 10,14 &  40,0 & 30,14 &  60,0 & 50,14 \\
22 &     &     &      &      & 12,8 &  14,4 &  34,8 &  36,4 &  56,8 &  58,4 \\
24 &     &     &      &      & 6,12 &   8,8 & 30,12 &  32,8 & 54,12 &  56,8 \\
26 &     &     &      &      & 2,12 &   4,8 & 28,12 &  30,8 & 54,12 &  56,8 \\
28 &     &     &      &      &  0,8 &   2,4 &  28,8 &  30,4 &  56,8 &  58,4 \\
30 &     &     &      &      &  0,0 &       &  30,0 & 18,18 &  60,0 & 48,18 \\
32 &     &     &      &      &      &       & 20,10 &  22,6 & 52,10 &  54,6 \\
34 &     &     &      &      &      &       & 12,16 & 14,12 & 46,16 & 48,12 \\ 
36 &     &     &      &      &      &       &  6,18 &  8,14 & 42,18 & 44,14 \\
38 &     &     &      &      &      &       &  2,16 &  4,12 & 40,16 & 42,12 \\
40 &     &     &      &      &      &       &  0,10 &   2,6 & 40,10 &  42,6 \\
42 &     &     &      &      &      &       &   0,0 &       &  42,0 & 28,22 \\
44 &     &     &      &      &      &       &       &       & 30,12 &  32,8 \\
46 &     &     &      &      &      &       &       &       & 20,20 & 22,16 \\
48 &     &     &      &      &      &       &       &       & 12,24 & 14,20 \\
50 &     &     &      &      &      &       &       &       &  6,24 &  8,20 \\
52 &     &     &      &      &      &       &       &       &  2,20 &  4,16 \\
54 &     &     &      &      &      &       &       &       &  0,12 &   2,8 \\
56 &     &     &      &      &      &       &       &       &   0,0 &       \\

\hline
\end{tabular}

\end{table*}


\begin{table}

\caption{
The valley of stability, obtained from Green's formula \cite{Green1955}. See Sec. \ref{valley} for further discussion.  
}
\begin{tabular}{ r r r r r r r r r r r r r r r    }
\hline
    &    &    &    &    &    &    &    &    &    &    &    &    &    &    \\
$Z$ & 30 & 32 & 34 & 36 & 38 & 40 & 42 & 44 & 46 & 48 & 50 & 52 & 54 & 56 \\
$N$ & 38 & 40 & 42 & 46 & 48 & 52 & 54 & 58 & 62 & 64 & 68 & 70 & 74 & 78 \\
    &    &    &    &    &    &    &    &    &    &    &    &    &    &    \\
$Z$ & 58 & 60 & 62 & 64 & 66 & 68 &  70 &  72 &  74 &  76 &  78 &  80 &  82 &    \\
$N$ & 80 & 84 & 88 & 92 & 84 & 98 & 102 & 106 & 110 & 114 & 118 & 120 & 124 &    \\
 &    &    &    &    &    &    &    &    &    &    &    &    &    &    \\
\hline
\end{tabular}

\end{table}


\begin{table}

\caption{
Nuclei along the valley of stability, obtained from Green's formula \cite{Green1955}, for which empirical values of $\gamma$ are available. The empirical values of the ratios $R_{4/2}$ and $R$ are obtained from Eqs. (\ref{R42}) and (\ref{R}) respectively, using data taken from the ENSDF database \cite{ensdf}, while the values $\gamma_R$ are obtained from Eq. (\ref{g}). The empirical values extracted in Ref. \cite{Lawrie2025} using the method involving only two $E2$ matrix elements are also shown, labeled by $\gamma_{TR}$, while the empirical values extracted \cite{Lawrie2025} through the Kumar-Cline method 
\cite{Kumar1972,Cline1986} are labeled by $\gamma_{KC}$. See Sec. \ref{valley} for further discussion. } 

\begin{tabular}{ r r r r r r }

\hline 
nucleus & $R_{4/2}$ & $R$  & $\gamma_R$ & $\gamma_{TR}$ & $\gamma_{KC}$ \\ 
\hline

\isotope[156][64]{Gd}$_{ 92}$ & 3.239 & 12.972 & 11.05 & 7.3(9)   &       \\ 
\isotope[160][66]{Dy}$_{ 94}$ & 3.270 & 11.133 & 11.90 &          &       \\
\isotope[166][68]{Er}$_{ 98}$ & 3.289 &  9.753 & 12.68 & 9.9(5)   & 18(3) \\
\isotope[172][70]{Yb}$_{102}$ & 3.305 & 18.616 &  9.27 & 5.0(7)   &  6(6) \\
\isotope[178][72]{Hf}$_{106}$ & 3.291 & 12.606 & 11.20 &          &       \\
\isotope[184][74]{ W}$_{110}$ & 3.273 &  8.122 & 13.84 & 11.4(3)  & 12(3) \\
\isotope[190][76]{Os}$_{114}$ & 2.934 &  2.988 & 22.28 & 23.3(13) & 25(2) \\
\isotope[196][78]{Pt}$_{118}$ & 2.465 &  1.936 &       & 38.8(11) &       \\

\hline
\end{tabular}

\end{table}


\begin{table*}

\caption{Single-particle states $(n_z, n_x, n_y)$ within the algebra U($(\eta+1)(\eta+2)/2$), where $n_z$, $n_x$, $n_y$ is the number of oscillator quanta along the $z$, $x$, and $y$ axes. Each line is characterized by the number $r$  of quanta removed from the $z$-axis \cite{Kota2020}. $n_r$ indicates the number of particles occupying the states within each line, while $N_r$ is the cumulative number of particles occupying 
all states up to the given line included. See Appendix C for further discussion.}  

\begin{tabular}{ r r r r r r r r r r r }

\hline 
$r$ &                 &                    &                   &                   &                   &                   &          &     & $n_r$ & $N_r$\\
\hline

0 & ($\eta$, 0, 0)    &                    &                   &                   &                   &                   &                   & &  2 &  2 \\
1 & ($\eta-1$, 1 , 0) &  ($\eta-1$, 0 , 1) &                   &                   &                   &                   &                   & &  4 &  6 \\
2 & ($\eta-2$, 2 , 0) &  ($\eta-2$, 1 , 1) & ($\eta-2$, 0 , 2) &                   &                   &                   &                   & &  6 & 12 \\
3 & ($\eta-3$, 3 , 0) &  ($\eta-3$, 2 , 1) & ($\eta-3$, 1 , 2) & ($\eta-3$, 0 , 3) &                   &                   &                   & &  8 & 20 \\
4 & ($\eta-4$, 4 , 0) &  ($\eta-4$, 3 , 1) & ($\eta-4$, 2 , 2) & ($\eta-4$, 1 , 3) & ($\eta-4$, 0 , 4) &                   &                   & & 10 & 30 \\
5 & ($\eta-5$, 5 , 0) &  ($\eta-5$, 4 , 1) & ($\eta-5$, 3 , 2) & ($\eta-5$, 2 , 3) & ($\eta-5$, 1 , 4) & ($\eta-5$, 0 , 5) &                   & & 12 & 42 \\
6 & ($\eta-6$, 6 , 0) &  ($\eta-6$, 5 , 1) & ($\eta-6$, 4 , 2) & ($\eta-6$, 3 , 3) & ($\eta-6$, 2 , 4) & ($\eta-6$, 1 , 5) & ($\eta-6$, 0 , 6) & & 14 & 56 \\
7 & ($\eta-7$, 7 , 0) &  ($\eta-7$, 6 , 1) & ($\eta-7$, 5 , 2) & ($\eta-7$, 4 , 3) & ($\eta-7$, 3 , 4) & ($\eta-7$, 2 , 5) & ($\eta-7$, 1 , 6) 
& ($\eta-7$, 0 , 7) & 16 & 72 \\
\hline
\end{tabular}

\end{table*}


\begin{figure} [htb]

    \includegraphics[width=75mm]{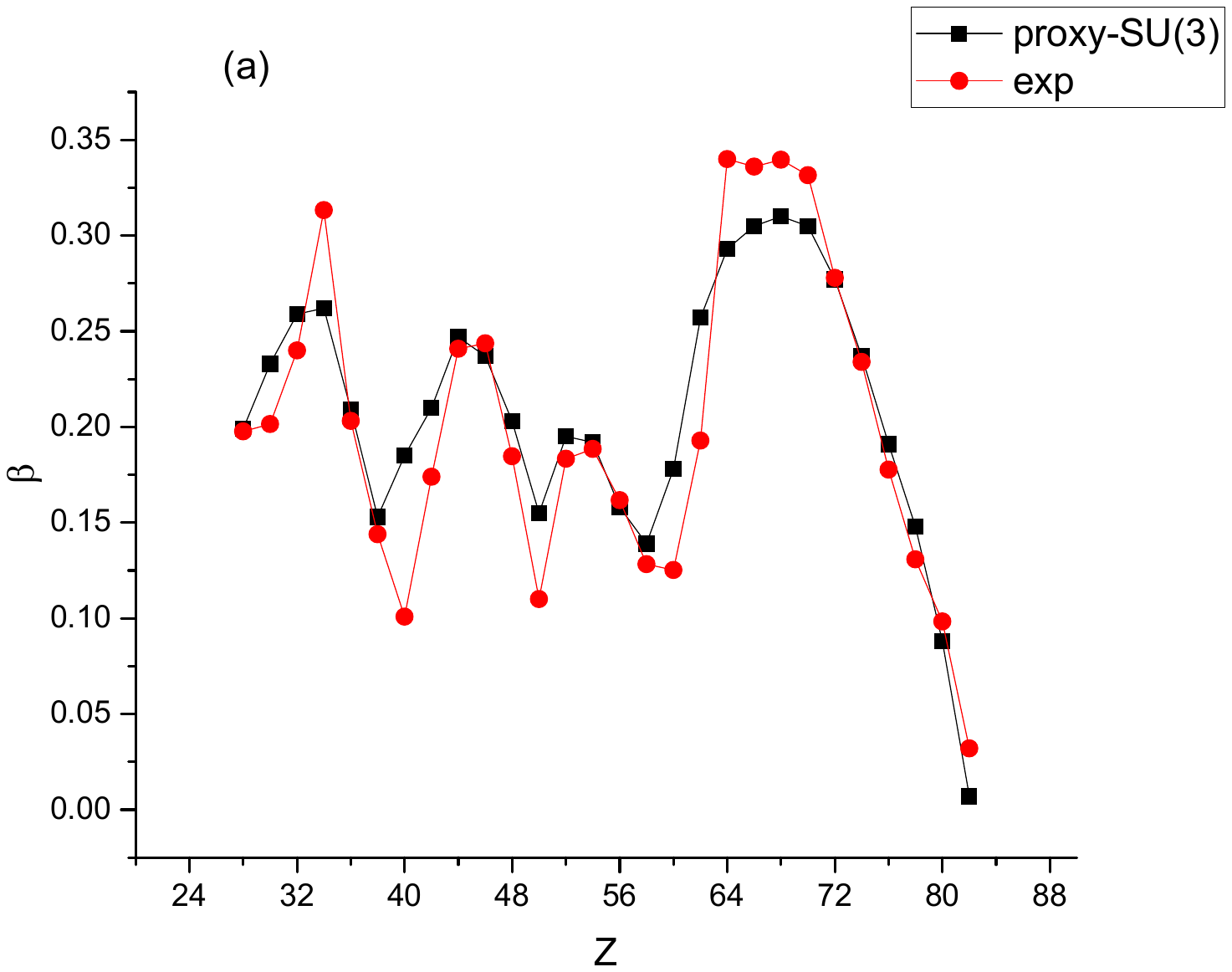} 
    \includegraphics[width=75mm]{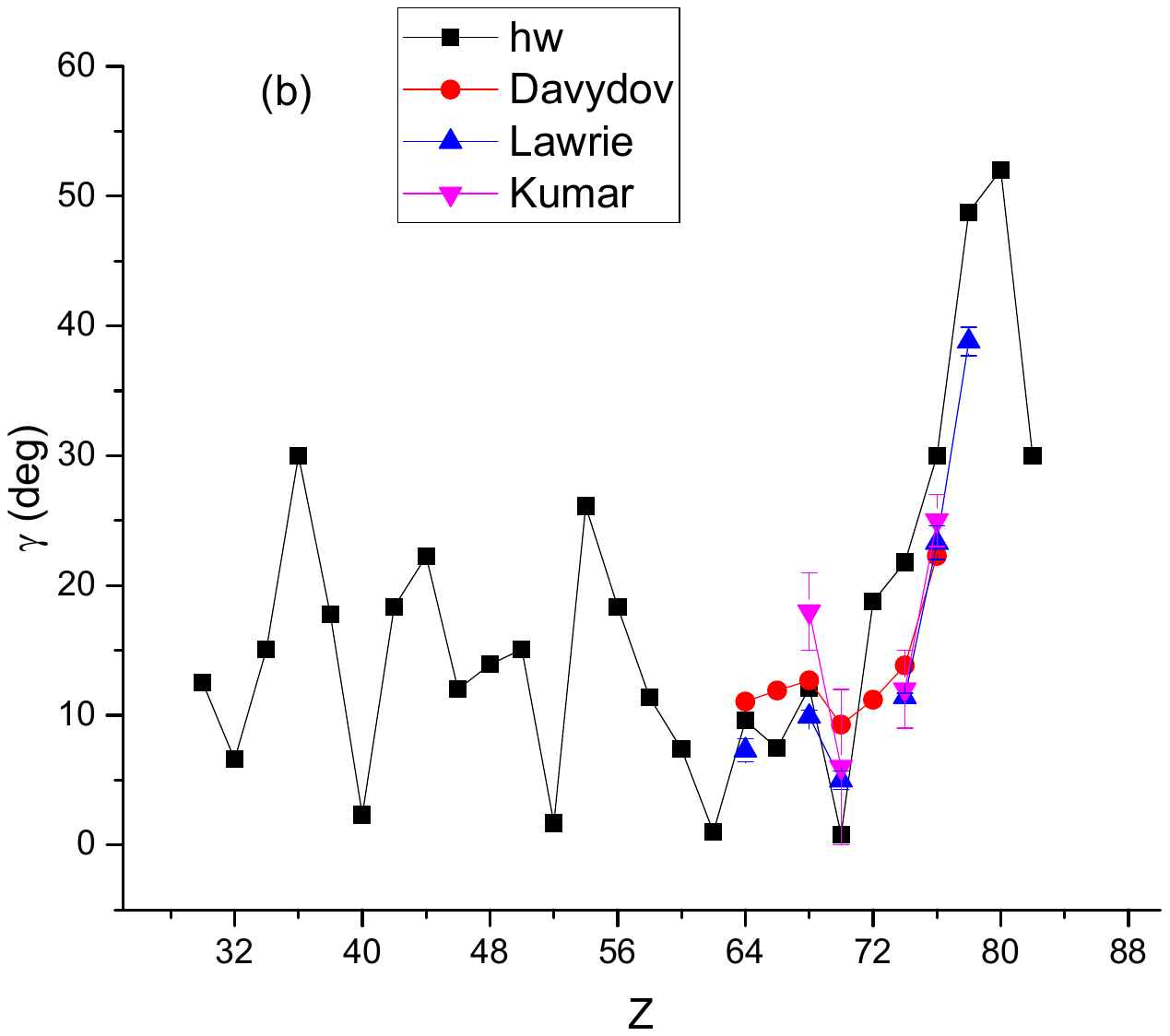} 
    \includegraphics[width=75mm]{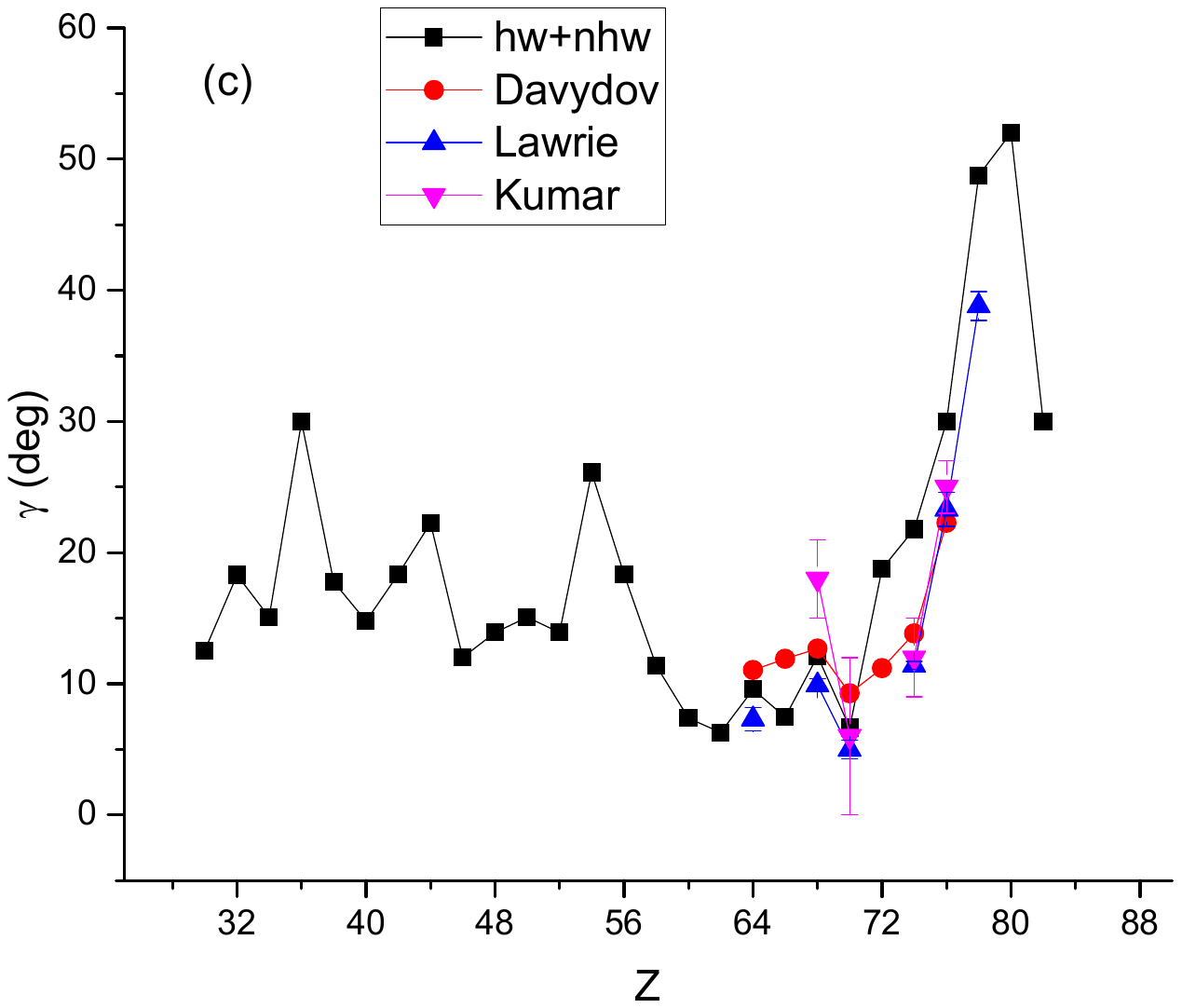} 
     
    \caption{(a) The parameter-free predictions for the collective variable $\beta$ along the valley of stability of Table VII \cite{Meng2022,Green1955}, obtained with the hw irrep of proxy-SU(3), are compared to the empirical values taken from Ref. \cite{Pritychenko2016}.
     (b) Parameter-free predictions for the collective variable $\gamma$ along the valley of stability, obtained with the hw irrep of proxy-SU(3) (labeled by hw), are compared to empirical values 
obtained through the Davydov model \cite{Davydov1958,Davydov1959} (labeled by Davydov), the method of Ref. \cite{Lawrie2025} (labeled by Lawrie), and the method of Kumar-Cline \cite{Kumar1972,Cline1986} (labeled by Kumar). 
(c) Same as (b), but with parameter-free predictions for the collective variable $\gamma$ obtained after mixing the hw and nhw irreps of proxy-SU(3) (labeled by hw+nhw). All empirical values of $\gamma$ are listed in Table VIII. See Sec. \ref{valley} for further discussion. } 
 
\end{figure}


\begin{figure} [htb]

    \includegraphics[width=75mm]{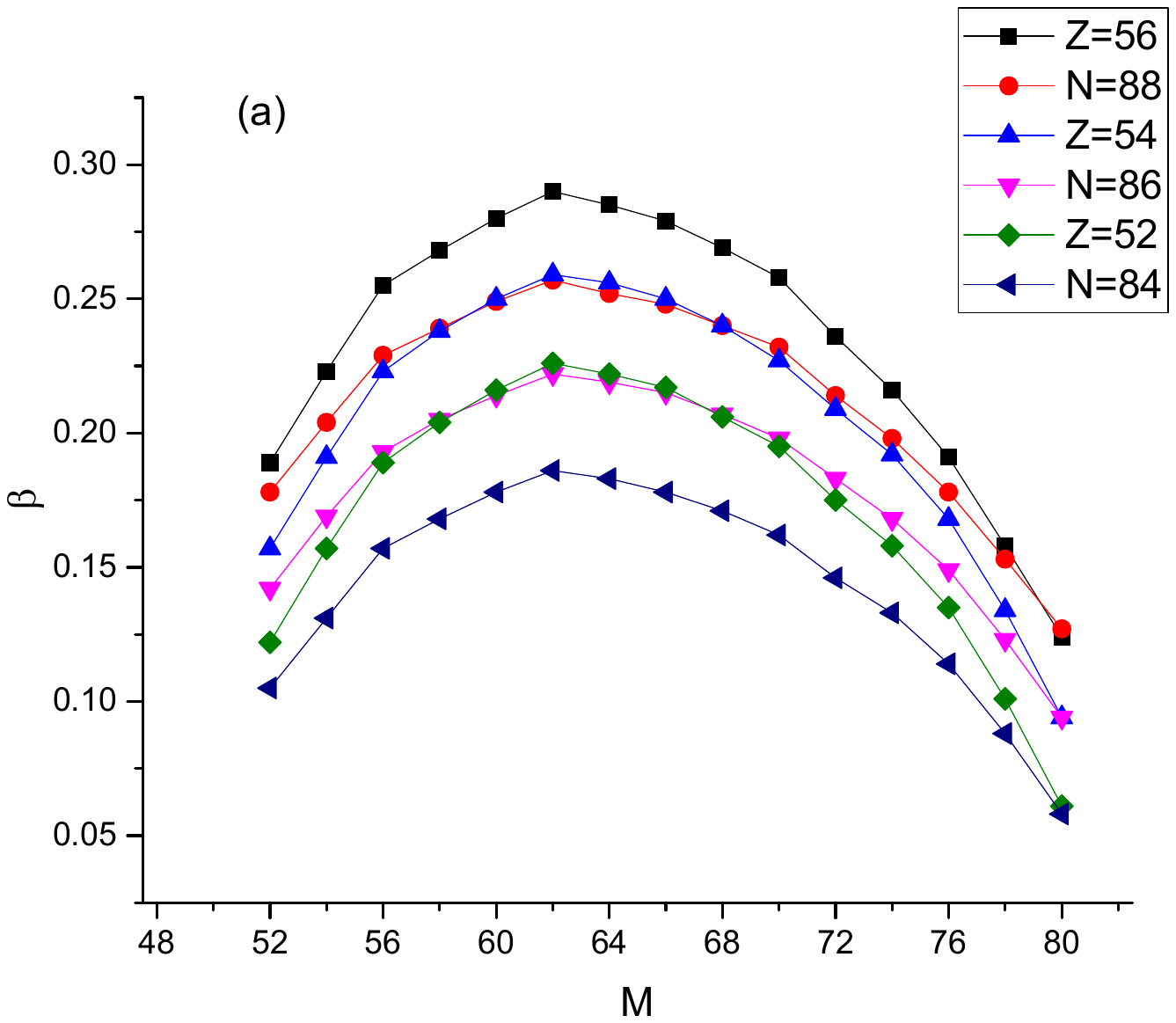} 
    \includegraphics[width=75mm]{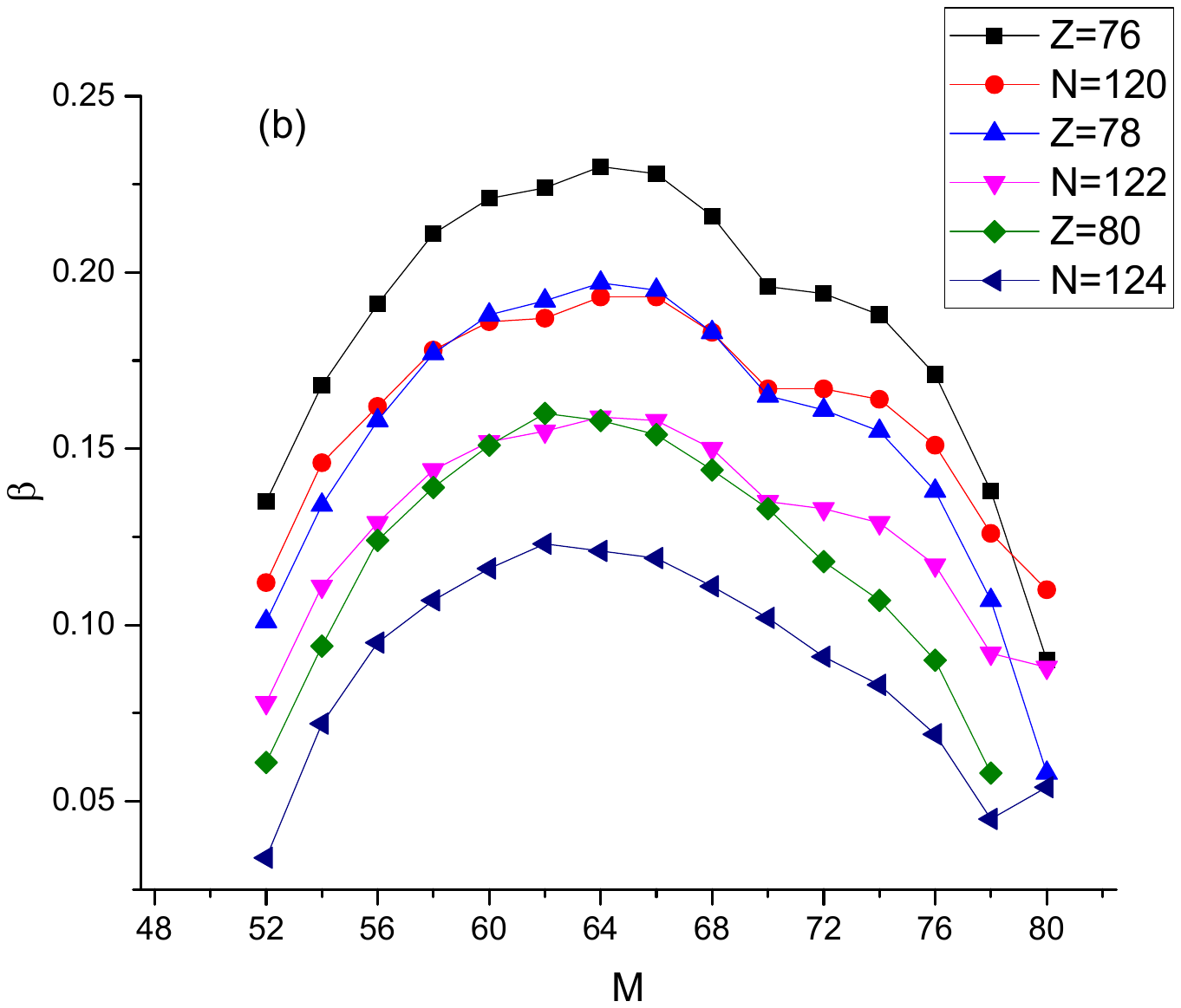}

    \caption{(a) The parameter-free predictions for the collective variable $\beta$ obtained with the hw irrep of proxy-SU(3) for $Z=52$, 54, 56 (in which case $M$ stands for the neutron number $N$) and $N=84$, 86, 88 (in which case $M$ stands for the proton number $Z$). 
b) Same for $Z=80$, 78, 76 and $N=124$, 122, 120. See Sec. \ref{mirror} for further discussion.}

\end{figure}


\begin{figure} [htb]

    \includegraphics[width=75mm]{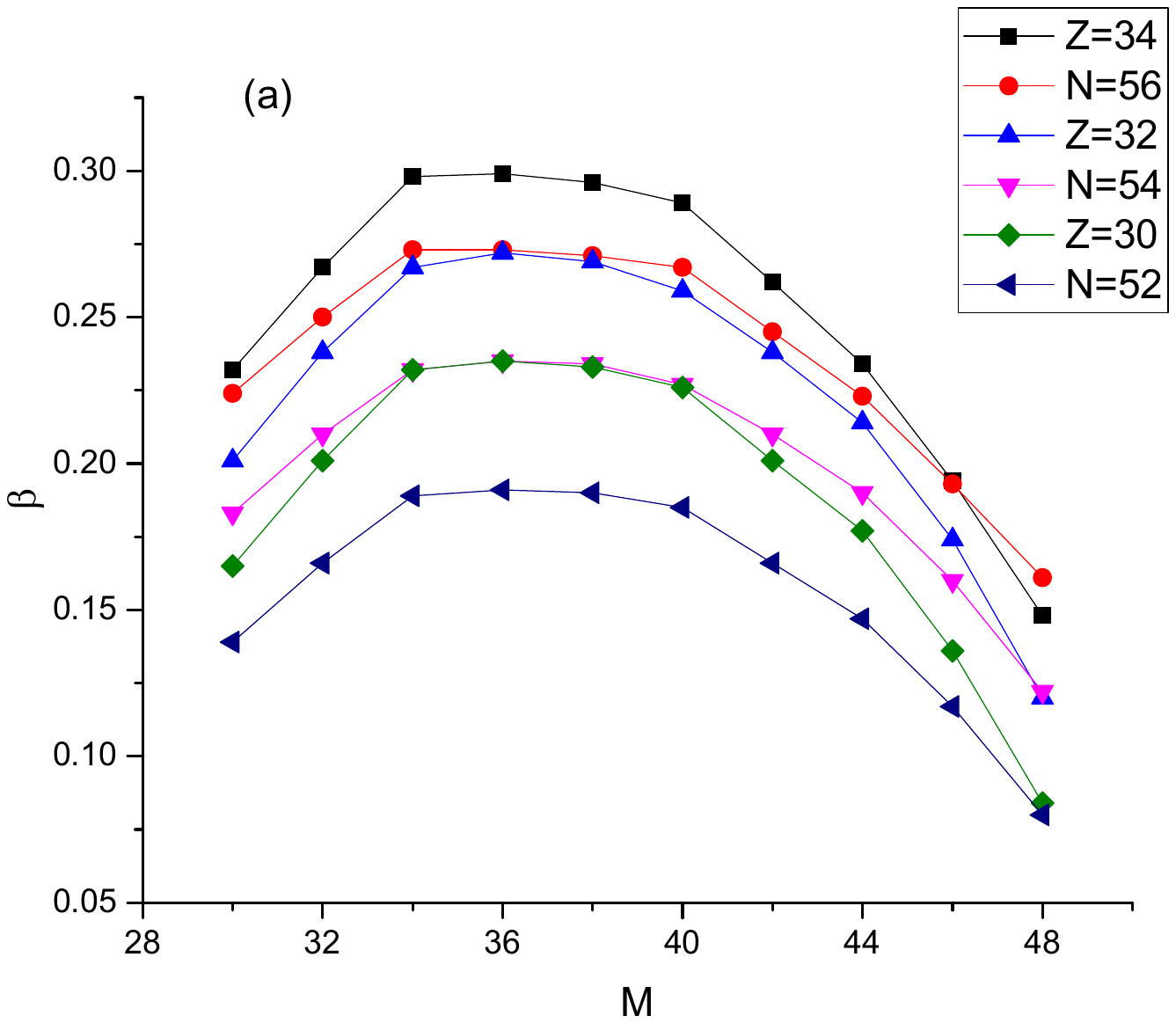} 
    \includegraphics[width=75mm]{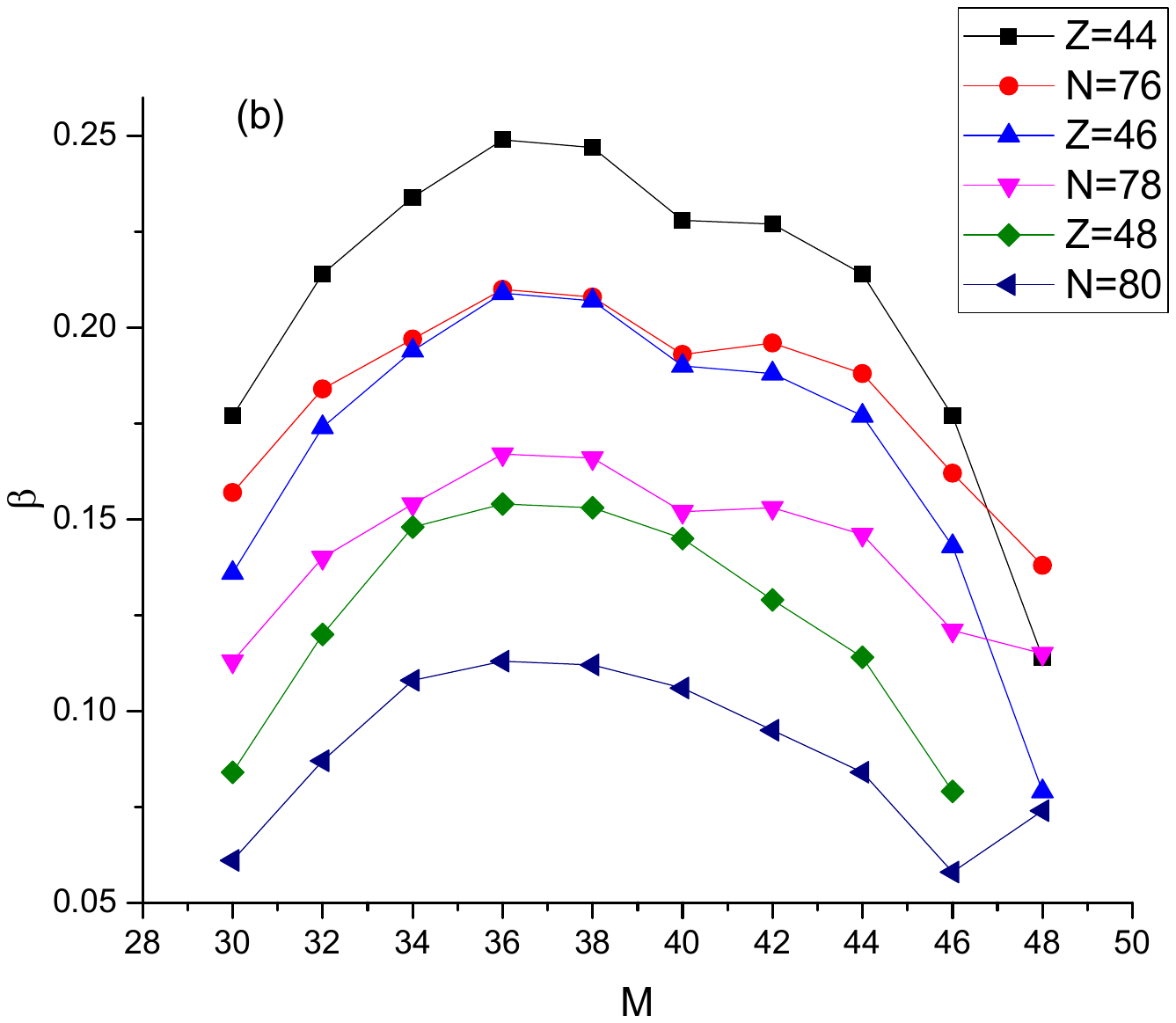}

  \caption{(a) The parameter-free predictions for the collective variable $\beta$ obtained with the hw irrep of proxy-SU(3) for $Z=30$, 32, 34 (in which case $M$ stands for the neutron number $N$) and $N=52$, 54, 56 (in which case $M$ stands for the proton number $Z$). b) Same for $Z=48$, 46, 44 and $N=80$, 78, 76. See Sec. \ref{mirror} for further discussion.}

\end{figure}

\end{document}